\documentclass{llncs}

\usepackage{graphicx}
\usepackage{mathptmx}
\usepackage{xurl}
\usepackage{mathptm}
\usepackage{amsmath}
\usepackage{amssymb}
\usepackage{amsfonts}
\usepackage{latexsym}
\usepackage{stmaryrd}
\usepackage{wrapfig}
\usepackage[table]{xcolor}
\usepackage{listings}
\usepackage{multirow}
\usepackage[ruled,vlined,linesnumbered]{algorithm2e}
\SetKwBlock{Repeat}{repeat}{}
\usepackage{caption}
\usepackage{thmtools} 
\usepackage{thm-restate}
\usepackage[capitalise]{cleveref}
\usepackage{enumerate}
\usepackage{pgf}
\usepackage{tikz}
\usetikzlibrary{automata,positioning,trees,shapes,petri,arrows,snakes,backgrounds,calc,fit,arrows.meta,shadows,math,decorations.markings}

\pgfdeclarelayer{mybackground}
\pgfdeclarelayer{background}
\pgfdeclarelayer{foreground}
\pgfsetlayers{background,mybackground,foreground,main}

\tikzstyle{background rectangle}=
[rounded corners,bottom color=green!10,top color=green!10,inner frame sep=0pt]
\tikzstyle{fitnode}=[draw=black, rounded corners,line width=0.5pt,dotted]
\tikzstyle{statenode}=
[circle,fill=blue!10,draw,inner sep=0.1pt, minimum size=3.2mm,text=blue!50!black,font=\footnotesize]
\tikzstyle{sstatenode}=
[circle,fill=blue!10,draw,inner sep=0.1pt, minimum size=7mm,text=blue!50!black,font=\footnotesize]
\tikzstyle{lblnode}=
[rounded corners,fill=red!10,inner sep=2pt,font=\footnotesize]
\tikzstyle{autedge}=[->,line width=0.5pt]
\tikzstyle{labelnode}=[text=black,align=left,sloped,above=-2pt,font=\footnotesize]
\tikzstyle{commentnode}=
[ellipse callout,draw,fill=green!10,inner sep=0.5pt, font=\footnotesize]
\tikzstyle{blocknode}=
[rectangle,rounded corners,draw,minimum width=50pt,minimum height=25pt,line width=2pt,align=center]
\tikzstyle{dotnode}=
[circle,fill=black,minimum size=6pt,inner sep=0pt]
\tikzstyle{cndnode}=
[kite,kite upper vertex angle=120,,kite lower vertex angle=120,draw,line width=2pt]
\tikzstyle{storenode}=
[ellipse,draw,line width=2pt,align=center]
\tikzstyle{blockedge}=[line width=2pt,->,>=stealth]
\tikzstyle{storeedge}=[dotted,line width=2pt,->,>=stealth]

\usepackage[colorinlistoftodos]{todonotes}

\title{Chain-Free String Constraints}
\subtitle{Technical report\footnote{This is an extended version of the work \cite{atva19} published at ATVA'19.}}

\author{Parosh Aziz Abdulla\inst{1} \and
Mohamed Faouzi Atig \inst{1} \and
   Bui Phi Diep \inst{1} \and
   Luk\'{a}\v{s} Hol\'{i}k\inst{2} \and
   Petr~Jank\r{u}\inst{2}}
\authorrunning{P. A. Abdulla et al.}
 \institute{Uppsala University, Sweden \\ \email{\{parosh,mohamed\_faouzi.atig,bui.phi-diep\}@it.uu.se} \and Brno University of Technology, Czech Republic \\ \email{\{holik,ijanku\}@fit.vutbr.cz}}

\begin{document}
\pagestyle{plain}
\maketitle              % typeset the header of the contribution

\begin{abstract}
We address the satisfiability problem for string constraints  that combine
relational constraints represented by transducers,
word equations,
and string length constraints. This problem is undecidable in general. Therefore,  we propose a  new decidable fragment of string constraints, called weakly chaining string constraints, for which we show that  the satisfiability problem is decidable. This fragment  pushes the borders of decidability of string constraints by generalising the existing straight-line as well as the acyclic fragment of the string logic.
We have developed a prototype  implementation of our new decision procedure, and integrated it into  in an existing framework that uses CEGAR with under-approximation of string constraints based on flattening. Our experimental results show the competitiveness and accuracy of the new framework.
%\keywords{String Constraints \and Satisfiability modulo theories \and Program verification}
\end{abstract}


%end{highlights}

%\begin{keyword}
%String-solving \sep string constraints \sep string program verification \sep web application vulnerability \sep automata \sep transducers \sep decidability



%\end{keyword}

%\end{frontmatter}

%\linenumbers


\newcommand{\formalmodel}[1]{{\sc #1}}
\newcommand{\FA}{{\formalmodel{FA}}}
\newcommand{\CFG}{{\formalmodel{CFG}}}
\newcommand{\RE}{{\formalmodel{RE}}}
\newcommand{\varedges}{\mathtt{var}}
\newcommand{\reledges}{\mathtt{rel}}
\newcommand{\nodevar}{\mathtt{var}}
\newcommand{\edgecons}{\mathtt{con}}
\newcommand{\defeq}{::=}
\newcommand{\formula}{\Psi}
\newcommand{\strterm}{\term_{\mathit{str}}}
\newcommand{\arterm}{\term_{\mathit{ar}}}
\newcommand{\memconstrof}[2]{#2(#1)}
\newcommand{\relconstrof}[3]{#3(#1,#2)}
\newcommand\posof[2]{(#1,#2)}
\newcommand\pos{p}

\newcommand{\Trau}[0]{{\sc Trau }}
\newcommand{\Traunospace}[0]{{\sc Trau}}
\newcommand{\Trauplus}[0]{{\sc Trau+}}
\newcommand{\Trauplusspace}[0]{{\sc Trau+ }}
\newcommand{\set}[1]{\left\{#1\right\}}
\newcommand{\setcomp}[2]{\left\{{#1}\mid {#2}\right\}}
\newcommand{\tuple}[1]{\left\langle#1\right\rangle}
\newcommand\fun{f}
\newcommand\gfun{g}

\newcommand{\ii}{i}
\newcommand{\jj}{j}
\newcommand{\kk}{k}
\newcommand{\mm}{m}
\newcommand{\nn}{n}
\newcommand{\pp}{p}

\newcommand{\aset}{A}
\newcommand{\sizeof}[1]{\left|#1\right|}
\newcommand{\intgrvarset}{V}
\newcommand{\intgrvar}{v}
\newcommand{\term}{t}
\newcommand{\termsover}[1]{{\tt Terms}\left(#1\right)}
\newcommand{\val}{\theta}
\newcommand{\valof}[1]{\val\left(#1\right)}
\newcommand{\pvalof}[1]{\val'\left(#1\right)}

\newcommand{\varset}{{\mathbb X}}
\newcommand{\nvarset}{{\mathbb U}}
\newcommand{\xvar}{x}
\newcommand{\yvar}{y}
\newcommand{\zvar}{z}
\newcommand{\eqn}{e}

\newcommand{\sanity}{{\tt Sanity}}
\newcommand{\sanityof}[2]{\sanity\left(#1\right)\left(#2\right)}
\newcommand{\genericof}[3]{{\tt Gen}\left(#1,#2\right)\left(#3\right)}
\newcommand{\sanityoneof}[2]{{\tt Sanity}_1\left(#1\right)\left(#2\right)}
\newcommand{\sanitytwoof}[2]{{\tt Sanity}_2\left(#1\right)\left(#2\right)}
\newcommand{\productof}[3]{{\tt Product}\left(#1,#2\right)\left(#3\right)}
\newcommand{\constr}{\varphi}
\newcommand{\constrs}{\psi}

\newcommand{\bool}{{\mathbb B}}
\newcommand{\nat}{{\mathbb N}}
\newcommand{\intgrs}{{\mathbb Z}}
\newcommand{\alphabet}{\Sigma}
\newcommand{\alphabete}{\Sigma_\emptystr}
\newcommand{\alphabetof}[1]{\alphabet\left(#1\right)}
\newcommand{\valphabetof}[3]{\alphabet^{#1}(#2,#3)}
\newcommand{\fsymbol}{f}
\newcommand{\asymbol}{a}
\newcommand{\bsymbol}{b}
\newcommand{\parikhof}[1]{{\tt Parikh}\left(#1\right)}
\newcommand{\compparikhof}[1]{{\tt CompP}\left(#1\right)}
\newcommand{\occurrence}{\#}
\newcommand{\occurrenceof}[1]{\occurrence{#1}}
\newcommand{\soccurrenceof}[2]{\occurrenceof{#1}\left(#2\right)}
\newcommand{\connect}{\varoast}
\newcommand{\strsover}[1]{{#1}^*}
\newcommand{\str}{w}
\newcommand{\istr}{v}
\newcommand{\istrof}[1]{\istr\left(#1\right)}
\newcommand{\pistrof}[1]{\istr'\left(#1\right)}
\newcommand{\lengthof}[1]{|#1|}
\newcommand{\emptystr}{\epsilon}
\newcommand{\addempty}[1]{{#1}_\emptystr}
\newcommand{\saddempty}[1]{{#1}^*_\emptystr}
\newcommand{\param}{{\mathbb p}}
\newcommand{\qaram}{{\mathbb q}}
\newcommand{\pqaram}{\alpha}
\newcommand{\mparam}{\alpha}
%\newcomman{\statenumof}[2]{{\tt StateNum}\left(#1,#2\right)}
%\newcommand{\statesof}[2]{{\tt States}\left(#1,#2\right)}
\newcommand{\statenumof}[1]{{\overline{\mathcal S}}\left(#1\right)}
\newcommand{\statesof}[1]{{\mathcal S}\left(#1\right)}
\newcommand{\transnumof}[1]{{\tt TransNum}\left(#1\right)}
\newcommand{\entriesof}[1]{{\tt Entries}\left(#1\right)}
\newcommand{\nonentriesof}[1]{{\tt NonEntries}\left(#1\right)}
\newcommand{\lastentryof}[1]{{\tt LastEntry}\left(#1\right)}
\newcommand{\flatformulaof}[2]{\phi\left(#1\right)}
\newcommand{\flangof}[1]{{\mathbb F}\left({#1}\right)}
\newcommand{\iflangof}[2]{{\mathbb F}^{#1}\left({#2}\right)}
\newcommand{\plangof}[1]{{\mathbb P}\left({#1}\right)}
\newcommand{\iplangof}[2]{{\mathbb P}^{#1}\left({#2}\right)}
\newcommand{\glangof}[1]{{\mathbb G}\left({#1}\right)}
\newcommand{\iglangof}[2]{{\mathbb G}^{#1}\left({#2}\right)}
\newcommand{\flatteningof}[2]{{\mathbb L}\left({#2}\right)\left({#1}\right)}
\newcommand{\flatten}[2]{{\tt Flatten}\left(#1\right)\left(#2\right)}
\newcommand{\mkplus}[1]{{#1}^\oplus}
\newcommand{\convert}[3]{{\mathbb h}\left(#1,#2\right)\left(#3\right)}
\newcommand{\cconvert}[5]{{\mathbb h}\left(#1\right)\left(#2,#3\right)\left(#4,#5\right)}
\newcommand{\loopsuccof}[2]{{\tt LoopSucc}\left({#1}\right)\left({#2}\right)}
\newcommand{\simploopsuccof}[1]{{\tt LoopSucc}\left({#1}\right)}
\newcommand{\rtloopsuccof}[2]{{\tt LoopSucc}^*\left({#1}\right)\left({#2}\right)}
\newcommand{\tloopsuccof}[2]{{\tt LoopSucc}^+\left({#1}\right)\left({#2}\right)}
\newcommand{\entrysuccof}[2]{{\tt EntrySucc}\left({#1}\right)\left({#2}\right)}

\newcommand{\simpentrysuccof}[1]{{\tt EntrySucc}\left({#1}\right)}

\newcommand{\succof}[2]{{\tt succ}\left({#1}\right)\left({#2}\right)}
\newcommand{\rtsuccof}[2]{{\tt succ}^*\left({#1}\right)\left({#2}\right)}
\newcommand{\tsuccof}[2]{{\tt succ}^+\left({#1}\right)\left({#2}\right)}
\newcommand{\starsuccof}[1]{{\tt succ}^*\left({#1}\right)}
\newcommand{\sigof}[1]{{\tt sig}\left(#1\right)}
\newcommand{\automaton}{{\mathcal A}}
\newcommand{\gautomaton}{{\mathcal B}}
\newcommand{\gautomatonof}[1]{\gautomaton\left(#1\right)}
\newcommand{\transducer}{{\mathcal T}}
\newcommand{\transducerof}[1]{\transducer\left(#1\right)}
\newcommand{\stateset}{Q}
\newcommand{\state}{q}
\newcommand{\transitionset}{\Delta}
\newcommand{\initstate}{\state_{\it init}}
\newcommand{\acceptingstate}{\state_{\it acc}}
\newcommand{\automatontuple}{\tuple{\stateset,\transitionset,I,F}}
\newcommand{\pautomatontuple}{\tuple{\stateset',\alphabet',\transitionset',\initstate',\acceptingstate'}}
\newcommand{\transducertuple}{\tuple{\stateset,\alphabet,\transitionset,\initstate,\acceptingstate}}
\newcommand{\acceptingof}[2]{{\tt accepting}\left({#1},{#2}\right)}
\newcommand{\automatontuplee}{\tuple{\stateset,\alphabet,\transitionset_1,\initstate,\acceptingstate}}
\newcommand{\automatontupleee}{\tuple{\stateset,\alphabet,\transitionset_2,\initstate,\acceptingstate}}

\newcommand{\simp}[1]{{\tt simp}\left({#1}\right)}
\newcommand{\Decomp}[1]{{\tt DECOMP}\left({#1}\right)}
\newcommand{\decomp}[1]{{\tt decomp}\left({#1}\right)}
\newcommand{\lhsdef}[0]{{\tt lhs}}
\newcommand{\rhsdef}[0]{{\tt rhs}}
\newcommand{\lhsf}[1]{{\tt lhs}\left({#1}\right)}
\newcommand{\LHSF}[1]{{\tt LHS}\left({#1}\right)}
\newcommand{\rhsf}[1]{{\tt rhs}\left({#1}\right)}
\newcommand{\RHSF}[1]{{\tt RHS}\left({#1}\right)}

\newcommand{\re}{{\mathcal R}}
\newcommand{\lang}{L}
\newcommand{\ilang}{K}
\newcommand{\clangof}[2]{\llangle#2\rrangle_{#1}}
\newcommand{\union}{+}
\newcommand{\concat}{\circ}
\newcommand{\app}{\bullet}
\newcommand{\langof}[1]{{\mathcal L}({#1})}
\newcommand{\relof}[1]{{\mathcal R}({#1})}
\newcommand{\eval}[1]{\eta({#1})}
\newcommand{\denotationof}[1]{\left\llbracket{#1}\right\rrbracket}
\newcommand{\renaming}{{\mathcal R}}
\newcommand{\renamingof}[1]{\renaming\left(#1\right)}
\newcommand{\prenaming}[1]{\renaming^{#1}}
\newcommand{\prenamingof}[2]{\prenaming{#1}\left(#2\right)}
 
\newcommand{\cfg}{{\mathcal G}}
\newcommand{\nonterminalset}{N}
\newcommand{\nonterminal}{A}
\newcommand{\terminal}{a}
\newcommand{\terminalset}{T}
\newcommand{\symb}{X}
\newcommand{\productionset}{P}
\newcommand{\production}{p}
\newcommand{\rhs}{\alpha}
\newcommand{\startsymbol}{S}
\newcommand{\cfgtuple}{\tuple{\nonterminalset,\productionset,\startsymbol}}

\newcommand{\phasevar}{{\tt phase}}

\newcommand\assigned\leftarrow

\newcommand\overf{{\tt Over}}
\newcommand\fresh{{\tt Fresh}}
\newcommand\upward{{\tt Upward}}
\newcommand\splitf{{\tt Split}}

\newcommand\spacex{\text{ }}
\newcommand\smallspace{\, \, \, \,} 
\newcommand\largespace{\, \, \, \, \, \, \, \, \, \, \, \, } 
\newcommand{\abc}[0]{{\sc ABC }}
\newcommand{\abcT}[0]{{\sc ABC-}{$\mathcal T$} }
\newcommand{\cba}[0]{{\sc CBA }}
\newcommand{\bca}[0]{{\sc BCA }}
\newcommand{\bcaT}[0]{{\sc BCA-}{$\mathcal T$} }
\newcommand{\bcanospace}[0]{{\sc BCA}}
\newcommand{\abcd}[0]{{\sc ABCD }}
\newcommand{\lukasFrag}[0]{{\sc LK }} 
\newcommand{\lukasFragnospace}[0]{{\sc LK}} 
\newcommand{\lukasFragT}[0]{{\sc LK-}{$\mathcal T$}}   

\newcommand\covered{{\tt Covered}}
\newcommand\waiting{{\tt Waiting}}
\newcommand\result{{\tt Result}}
\newcommand\uapproxres{{\tt UAprxResult}}
\newcommand\uapprox{{\tt UAprx}}
\newcommand\uapproxof[2]{\uapprox\left(#1,#2\right)}
\newcommand\oapproxres{{\tt OAprxResult}}
\newcommand\oapprox{{\tt OAprx}}
\newcommand\oapproxof[1]{\oapprox\left(#1\right)}
\newcommand\wconstrs{{\mathbb W}}
\newcommand\return{{\tt return}}
\newcommand\returnof[1]{\return\left(#1\right)}
\newcommand\unsat{{\tt unsat}}
\newcommand\gen{{\tt GenPar}}
\newcommand\genof[1]{\gen\left(#1\right)}
\newcommand\qelim{{\tt QElim}}
\newcommand\qelimof[1]{\qelim\left(#1\right)}
\newcommand\smt{{\tt SMT}}
\newcommand\smtof[1]{\smt\left(#1\right)}
\newcommand\nconstr\varrho
\newcommand\pconstr\rho
\newcommand\fvars{{\tt FV}}
\newcommand\fvarsof[1]{\fvars\left(#1\right)}
\newcommand\atoms{{\tt Atoms}}
\newcommand\atomsof[1]{\atoms\left(#1\right)}
\newcommand\true{{\tt true}}
\newcommand\pof[1]{{#1}^\bullet}
\newcommand\nvarof[3]{\lessdot #1,#2,#3 \gtrdot}
\newcommand\genistrof[1]{{\tt GenIStr}\left(#1\right)}
\newcommand\addvar\ostar
\newcommand\restrictof[2]{\left[#1\right]_{#2}}
\newcommand\getstr{{\tt GetS}}
\newcommand\getstrof[1]{\getstr\left(#1\right)}

\newcommand\yfc[1]{{\color{blue}\ (YFC: #1) }}
\newcommand\parosh[1]{{\color{red}\ (Parosh: #1) }}
\newcommand\diep[1]{{\color{red}\ (Diep: #1) }}
\newcommand{\lukas}[1]{}
\newcommand\ahmed[1]{{\color{olive}\ (Ahmed: #1) }}
%\renewcommand{\lukas}[1]{}

%from petrDraft
\newcommand{\ruleName}[1]{\textsc{#1}}

\newcommand{\init}{I}
\newcommand{\final}{F}
\newcommand{\afa}{\mathcal{A}}
\newcommand{\aft}{\mathcal{R}}
\newcommand{\states}{Q}
\newcommand{\transitions}{\Delta}
\newcommand{\aftrel}{\mathcal{R}}
\newcommand{\aftrels}{\mathcal{S}}
\newcommand{\afare}{P}

\newcommand{\map}{\mathtt{map}}
\newcommand{\rel}{R}
\newcommand{\var}{\mathtt{var}}
\newcommand{\lab}{\mathtt{label}}
\newcommand{\nodes}{N}
\newcommand{\edges}{E}
\newcommand{\splitrel}{\textsc{Rel-Split}}
\newcommand{\splitleft}{\textsc{Split}_\leftarrow}
\newcommand{\splitright}{\textsc{Split}_\rightarrow}
\newcommand{\leftsplit}{\Phi_L}
\newcommand{\rightsplit}{\Phi_R}

\newcommand{\modulo}{\mod}
\newcommand\clause{\Upsilon}
\newcommand\clauseone{\bar\Upsilon}
\newcommand{\drop}{\mathit{drop}}
\newcommand{\reminder}{{\mathsf{r}\!\clause}}

\section{Introduction}
The recent years have seen  many works dedicated to extensions of SMT solvers with new background theories that can lead to efficient analysis   of programs with high-level data types. A data type that has attracted a lot of attention is {\em string} (for instance \cite{Chen:2019,Chen:2018,Kiezun09hampi,fangyu:stranger,Zheng13z3str,trinh2014:s3,LiaEtAl-CAV-14,KauslerS14,string14,flatten17,BL16,Berzish2017Z3str3AS,sloth}). Strings are present in almost all programming and scripting languages. String solvers can be extremely useful in applications such as verification of string-manipulating programs \cite{string14} and analysis of security vulnerabilities of scripting languages (e.g., \cite{Saxena10:kaluza,saxena:flax,fangyu:stranger,BL16}).
The wide range of the commonly used primitives for manipulating strings in such languages requires string solvers to handle an expressive class of string logics. The most important features that a string solver have to model are {\em concatenation} (which is used to express assignments in programs), {\em transduction} (which can be  used to model sanitisation and replacement operations), and {\em string length} (which is used to constraint lengths of strings). 

It is well known that the satisfiability problem for the full class of string constraints with concatenation, transduction, and length constraints is undecidable in general \cite{morvan,Chen:2019} even for a simple formula of the form $\transducer(x,x)$ where $\transducer$ is a rational transducer and $x$ is a string variable. 
However, this  theoretical barrier did not prevent the development of numerous efficient solvers
such as Z3-str3 \cite{Berzish2017Z3str3AS}, Z3-str2~\cite{Zheng13z3str}, 
CVC4~\cite{LiaEtAl-CAV-14}, S3P~\cite{trinh2014:s3,Trinh2016},
 and  \Trau \cite{flatten17,trau18}. 
These tools implement semi-algorithms to handle a large variety of string constraints, but do not provide completeness  guarantees. 
%of termination. %of their decision procedures. 
Another direction of research is to  find meaningful and expressive subclasses of string logics for which the satisfiability problem is decidable. 
Such classes include the  acyclic fragment of Norn~\cite{norn15}, the solved form fragment \cite{Ganesh13decide}, and also the \mbox{straight-line fragment \cite{BL16,sloth,Chen:2018}.}

In this paper, we propose an approach  which is a mixture of the two above research directions, namely  finding decidable fragments and making use of it to develop  efficient  semi-algorithms. To that aim, we define the class of {\em chain-free} formulas which strictly subsumes the acyclic fragment of  Norn~\cite{norn15} as well as the straight-line fragment of
\cite{BL16,sloth,Chen:2018}, and thus extends the known border of decidability for string constraints. 
The extension is of a practical relevance. 
A straight-line constraint models a path through a string program in the single static assignment form,
but as soon as the program compares two initialised string variables, the string constraint falls out of the fragment. 
The acyclic restriction of Norn on the other hand  does not include transducer constraints %(although it might be extended to them) 
and does not allow multiple occurrences of a variable in a single string constraint (e.g. an equation of the form $xy = zz$).
Our chain-free fragment is liberal enough to accommodate constraints that share both these forbidden features (including  $xy = zz$). 
%

%% \newcommand{\code}[1]{{\tt #1}}
%% \newcommand{\bad}{\mathit{Danger}}
%% The following pseudo-PHP code \footnote{the original at \cite{script}} that prompts user to change his password is an example of a program that is chain free but does not fall into existing decidable fragments.
%% %\begin{minted}[fontsize=\footnotesize]{php}
%% {\small
%% \begin{verbatim}
%% $oldS = $database->real_escape_string($oldU);
%% $newS = $database->real_escape_string($newU);
%% $querry1 = "SELECT password FROM users WHERE userID=$user";
%% $pass = $database->query($querry1);
%% if($old == $pass]) 
%% 	querry2 = "UPDATE users SET password=$new WHERE userID=$user";
%%    $database->query(querry2);
%% \end{verbatim}
%% }
%% 
%% The user inputs the old password \code{oldU} and the new password \code{newU},
%% both are sanitized and assigned to \code{oldS} and \code{newS}, respectively, 
%% the sanitized old password is compared with the users password in the database,
%% and if they match, the new sanitized password is saved in the database.  
%% %
%% The sanitization is present to prevent SQL injection attack.
%% %
%% To verify that the sanitization works, we need to verify that the SQL queries \code{querry1} and \code{querry2} can never conform a dangerous regular pattern $\bad$.
%% %
%% That the second SQL query is safe is expressed (after some slicing) by the constraint 
%% $$
%% \code{oldS} = T(\code{oldU}) \land \code{newS} = T(\code{newS}) 
%% %\land \code{querr1} = w_1.\code{user} 
%% \land \code{old} = \code{pass}
%% \land \code{querry2} = "u.\code{new}.v.\code{user}
%% $$

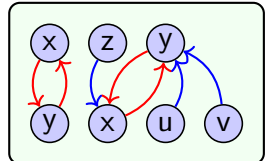
\begin{wrapfigure}[7]{r}{0.33\textwidth}
\centerline{
\begin{tikzpicture}[rotate=90,framed,
background rectangle/.style={thick,draw=black,rounded corners,fill = green!5}, 
scale=0.7, state/.style={fill=blue!20, draw, circle, semithick, minimum width=0.5cm, inner sep=0pt, text centered, font = \large \sffamily},transform shape]
  [auto,
  	->,>=stealth',shorten >=2pt,auto,node distance=2cm,on grid,semithick  	
	]

%\node[state] (x)							{\large \sffamily x};
%\node[state] (y)	[below right=1cm of x]	{\large \sffamily t2};
%\node[state] (z)	[above right=1cm of x]	{\large \sffamily z};

\matrix [row sep=0.55cm, column sep = 0.25cm]
%{	\node [state] (x) {x}; 	& \node [state] (y) {y};		 		\\ 
%	\node [state] (y1) {y}; & \node [state] (t1) {t};		 		\\ 
%	\node [state] (z1) {z}; & \node [state] (t2) {t};\\
%	 								& \node [state] (x1) {x};\\ 
%};
{	
	\node [state] (x) {x}; &\node [state] (z1) {z}; & \node [state] (y1) {y}; \\
	\node [state] (y) {y}; & \node [state] (x1) {x}; & \node [state] (u1) {u}; & \node [state] (v1) {v};\\
};

\path[-, thick, color=red,->] (x) edge			[bend right]		node{} 		 (y);	
\path[-, thick, color=red,->] (y) edge			[bend right]		node{}		 (x);	
\path[-, thick, color=red,->] (y1) edge		[bend right]		node{}		 (x1);	
\path[-, thick, color=red,->] (x1) edge		[bend right]		node{}		 (y1);	
\path[-, thick, color=blue,->] (z1) edge		[bend right]		node{}		 (x1);	
\path[-, thick, color=blue,->] (u1) edge		[bend right]		node{}		 (y1);	
\path[-, thick, color=blue,->] (v1) edge		[bend right]		node{}		 (y1);	
\end{tikzpicture}
}
\vspace{-0.3cm}
\caption{The splitting graph of $x =  z \cdot y \wedge  y = x \cdot u \cdot v$.}
\label{fig:graph}
\end{wrapfigure}

\newcommand{\code}[1]{{\tt #1}}
\newcommand{\bad}{\mathit{Bad}}
\newcommand{\good}{\mathit{Good}}
The following pseudo-PHP code
(a variation of a code at \cite{script}) 
that prompts a user to change his password is an example of a program that generates a chain-free constraint that is neither straight-line nor acyclic according to \cite{BL16,string14}.
%\begin{minted}[fontsize=\footnotesize]{php}
{\footnotesize 
\begin{verbatim}
$old=$database->real_escape_string($oldIn);
$new=$database->real_escape_string($newIn);
$pass=$database->query("SELECT password FROM users WHERE userID=".$user);
if($pass == $old) 
   if($new != $old)
      $query = "UPDATE users SET password=".$new." WHERE userID=".$user;
      $database->query($query);
\end{verbatim}
}
\noindent
The user inputs the old password \code{oldIn} and the new password \code{newIn},
both are sanitized and assigned to \code{old} and \code{new}, respectively.
The old sanitized password is compared with the value \code{pass} from the database, 
to authenticate the user, 
and then also with the new sanitized password, 
to ensure that a different password was chosen,
and finally saved in the database.
%\lukas{say also that pass is unrestricted since there can be anything in the database and that user is not user input so we don't care about the first SQL query?} 
%
The sanitization is present to prevent SQL injection.
To ensure that the sanitization works, 
we wish to verify that the SQL query \code{query} is safe, that is, it does not belong to a regular language $\bad$ of dangerous inputs.
This safety condition is expressed  by the constraint 
\begin{multline*}
%\hspace{-5mm}
\code{new} = \transducer(\code{newIn}) 
\land \code{old} = \transducer(\code{oldIn})
\land \code{pass} = \code{old}
\land \code{new} \neq \code{old}
\\
\land \code{query} = u.\code{new}.v.\code{user}
\land \code{query}\in\bad
\end{multline*}
The sanitization on lines 1 and 2 is modeled by the transducer $\transducer$,
and $u$ and $v$ are the constant strings from line 7. 
The constraints fall out from the straight-line due to the test $\code{new}\neq\code{old}$.
The main idea behind the chain-free fragment is to associate to the set of relational constraints  a \emph{splitting graph} where each node corresponds to an occurrence of a variable in the relational constraints of the formula (as shown in Figure \ref{fig:graph}).
An edge from an occurrence of $x$ to an occurrence of $y$ means that the source occurrence of $x$ appears in a relational constraint which has in the opposite side an occurrence of $y$ different from the target occurrence of $y$. 
The chain-free fragment prohibits loops in the graph, that we call \emph{chains}, such as those shown in red in Figure~\ref{fig:graph}.

\setlength{\intextsep}{0pt}%

Then, we identify  the so called \emph{weakly chaining} fragment which strictly extends the chain-free fragment by allowing  \emph{benign} chains.
Benign chains relate relational constraints where each left side contains only one variable, the constraints are all \emph{length preserving}, 
and all the nodes of the cycles appear exclusively on the left or exclusively on the right sides of the involved relational constraints
(as is the case in Figure \ref{fig:graph}).
Weakly chaining constraints may in practice arise from the checking that an encoding followed a decoding function is indeed the identity, i.e., satisfiability of constraints of the form $\transducer_{\text{enc}}(\transducer_{\text{dec}}(x)) = x$, discussed e.g. in \cite{Hu:2017}. For instance, in situations similar to the example above, one might like to verify that the sanitization  of  a password followed by the application of  a function supposed to invert the sanitization gives the original password.

Our decision procedure for the weakly chaining formulas proceeds in several steps.
The formula is  transformed to an equisatisfiable  chain-free formula, and then to an equisatisfiable concatenation-free formula in which the relational constraints are of the form $\transducer(x,y)$ where $x$ and $y$ are two string variables and $\transducer$ is a transducer/relational constraint. 
Finally, we provide a decision procedure of a chain and concatenation-free formulae.
The algorithm is  based on two  techniques.
First,  we show that the chain-free conjunction over relational constraints can be turned into a single equivalent transducer constraint (in a similar manner as in \cite{BFL13}).
Second, consistency of the resulting transducer constraint with the input length constraints is checked via the computation  of the Parikh image of the transducer.

To demonstrate  the usefulness of our approach, we have implemented our decision procedure in {\sc Sloth} \cite{sloth}, 
and then integrated it in the open-source  solver \Trau \cite{flatten17,trau18}.  \Trau is a string solver which  is based on  a Counter-Example Guided Abstraction Refinement (CEGAR) framework which contains both an under- and an over-approximation module. These two modules interact together in order to automatically  make these approximations more precise. We have implemented our decision procedure inside the over-approximation module which takes as an input a constraint and checks if it belongs to the weakly chaining fragment. If it is the case, then we use our decision procedure outlined above. Otherwise, we start  by  choosing a minimal set of  occurrences of  variables $\xvar$ that needs to be  replaced by  fresh ones such that the resulting constraint falls in our decidable fragment.  
We compare  our prototype implementation against four other state-of-the-art string solvers, namely Ostrich \cite{Chen:2019}, Z3-str3 \cite{Berzish2017Z3str3AS}, CVC4 \cite{LiaEtAl-CAV-14,CVC4}, and  {\sc Trau}  \cite{Trau}. For our comparison with Z3-str3, we use the version that is part of Z3 4.8.4.
Our experimental results show the competitiveness as as well as accuracy of the framework compared to the solver \Trau \cite{flatten17,trau18}. Furthermore, the experimental results show the competitiveness  and generality of our method compared to the existing techniques.  In summary, our main contributions are: $(1)$ a new decidable fragment of string constraints, called chain-free, which strictly generalises the existing straight-line as well as the acyclic fragment \cite{BL16,string14} and precisely characterises the decidability limitations of general relational/transducer constraints combined with concatenation,  $(2)$  a relaxation of the chain-free fragment, called  weakly chaining, which allows special chains with length preserving relational constraints, $(3)$  a decision procedures for checking the satisfiability problem of chain-free as well as weakly chaining string constraints, and $(4)$  a prototype with experimental
results that demonstrate the efficiency and generality of our technique on benchmarks from the literature as well as on new benchmarks.

\section{Preliminaries}
\label{sec:preliminaries}

%\medskip
%
%\noindent
\paragraph{Sets and strings.}
We use $\nat$, $\intgrs$ to denote the sets of natural numbers 
and integers, respectively.
A finite set $\alphabet$ of \emph{letters} is an \emph{alphabet},
a sequence of symbols $a_1 \cdots a_n$ from $\alphabet$ is a \emph{word} or a \emph{string} over $\alphabet$, 
with its \emph{length} $n$ denoted by $\lengthof\str$, 
$\emptystr$ is the \emph{empty word} with $\lengthof\emptystr = 0$, it is a neutral element with respect to string concatenation $\concat$, 
and $\strsover\alphabet$ is the set of all words
over $\alphabet$ including $\emptystr$.

%\medskip
%
%
%\noindent
\paragraph{Logic.}
Given a predicate formula, an occurrence of a predicate is \emph{positive} if it is under an even number of negations.
A formula is in \emph{disjunctive normal form} (DNF) if it is a disjunction of \emph{clauses} that are themselves conjunctions of (negated) predicates.
We write $\formula[x/\term]$ to denote the formula obtained by substituting in the formula $\formula$ each occurrence of the   variable $x$ by the term $\term$.

%\medskip
%
%
\paragraph{(Multi-tape)-Automata and transducers.}
A {\it Finite Automaton} (\FA) over an alphabet $\alphabet$
is a tuple $\automaton=\automatontuple$, where
$\stateset$ is a finite set of {\it states},
$\transitionset\subseteq\stateset\times\alphabete\times\stateset$ 
with $\alphabete = \alphabet\cup\{\epsilon\}$
is a set of {\it transitions},  and 
$I\subseteq\stateset$ (resp. $F\subseteq\stateset$ ) are the {\it initial} (resp. {\it accepting}) states.
$\automaton$ accepts a word $w$ iff there is a sequence 
$q_0 a_1 q_1 a_2 \cdots a_n q_n$ such that $(q_{i-1}, a_i, q_i)\in\transitionset$ for all $1\leq i \leq n$, 
$q_0\in I$, $q_n\in F$, and 
$w = a_1\concat \cdots \concat a_n$. 
The \emph{language} of $\automaton$, denoted $\langof\automaton$, is the set all accepted words. 

Given $n \in \mathbb{N}$, a \emph{$n$-tape automaton} $\transducer$ is an automaton over the alphabet $(\alphabet_\emptystr)^n$. It \emph{recognizes} the relation  $\relof{\transducer}\subseteq (\alphabet^*)^n$ that contains vectors of words $(w_1,w_2, \ldots, w_n)$ for which there is $(a_{(1,1)},a_{(2,1)},\ldots,a_{(n,1)} )\cdots(a_{(1,m)},a_{(2,m)},\ldots,a_{(n,m)})\in\langof\transducer$ with $w_i = a_{(i,1)} \concat \cdots\concat a_{(i,m)}$ for all $i \in \{1,\ldots,n\}$. 
A $n$-tape automaton $\transducer$ is said to be {\em length-preserving} if its transition relation $\transitionset\subseteq\stateset\times \alphabet^n \times\stateset$. A \emph{transducer} is  a $2$-tape automaton.
 
Let us recall some closure properties of multi-tape automata. First, the class of length-preserving $n$-tape automata is closed under union, intersection, and complement (by standard automata constructions). General $n$-tape automata are still closed under union, but not under complementation nor intersection. 
%However, the class  of length-preserving  multi-tape automata is closed under intersection.  
Multi-tape automata are closed under \emph{composition}:
  Let $\transducer$ and $\transducer'$ be two multi-tape automata of dimension $n$ and $m$, respectively, and let $i \in \{1,\ldots,n\}$ and $j \in \{1,\ldots,m\}$ be two indices. An $(n+m-1)$-tape automaton $\transducer \wedge_{(i,j)} \transducer'$ which accepts the set of words $(w_1,\ldots,w_n,u_1, \ldots, u_{j-1},$ $u_{j+1},\ldots,u_m)$ if and only if $(w_1,\ldots,w_n) \in \relof{\transducer}$ and $(u_1,\ldots, u_{j-1},w_i,u_{j+1},\ldots,u_m) \in \relof{\transducer'}$ is then constructed as a product of the two automata that synchronises over their $i$th and $j$th track, respectively. Multi-tape automata are also closed under \emph{permutation}: Given a permutation $\sigma\,:\, \{1,\ldots,n\} \rightarrow \{1,\ldots,n\}$ and a $n$-tape automaton $\transducer$, an $n$-tape automaton $\sigma(\transducer)$ such that $\relof{\sigma(\transducer)}=\{(w_{\sigma(1)}, \ldots,w_{\sigma(n)}) \, |\, (w_1,w_2, \ldots,w_n) \in\relof{\transducer}\}$ is obtained by replacing symbols $(a_1,\ldots,a_n)\in\Sigma^n$ on transitions by their permutations $(\sigma(a_1),\ldots,\sigma(a_n))$. Further, a \emph{cylindrification} of an $n$-tape automaton $\transducer$ for a natural number $k\geq n$ is a $k$-tape automaton s. t. $(w_1,\ldots,w_k)\in \relof{\transducer'}$ if and only if $(w_1,\ldots,w_n)\in \relof{\transducer}$, and it is obtained by replacing each transition over a symbols $(a_1,\ldots,a_n)\in\Sigma^n$ by a set of transitions $\{(a_1,\ldots, a_k)\mid a_{n+1}\,\ldots,a_{k}\in\Sigma\}$.
%
%\lukas{moved here from the proof of Lemma XY:}
Finally, the  \emph{concatenation} of $n$-tape automata $\transducer$ and $\transducer'$, an $n$-tape automaton accepting the relation $\relof{\transducer}.\relof{\transducer'} = \{(x.y,x'.y')\mid (x,y)\in\relof{\transducer}\land (x',y')\in\relof{\transducer'}\}$, can be constructed e.g. by connecting finals states of $\transducer$ to initial states of $\transducer'$ with $\epsilon^n$-transitions.

%\newpage
\section{String Constraints}
\label{constraints:section}

\begin{wrapfigure}[7]{r}{0.5\textwidth}
\begin{tikzpicture}[framed,background rectangle/.style={thick,draw=black,rounded corners,fill = green!5}, scale=1, transform shape]
%\small
\node (species1) {
		%\begin{tabular}{r c l r} 
		%$\constrs$ 		& ::= & 		$\constr$ $|$ $\constrs \wedge \constrs$\\
		%$\constr$ 		& ::= & 		$\constr_{s}$ $|$ $\constr_{i}$ $|$ $\constr_{t}$ $|$ $\constr_{g}$\\
		%$\constr_{s}$ 	& ::= & 		$\term_{s} = \term_{s}$ $|$ $\term_{s} \neq \term_{s}$ \\
		%$\constr_{t}$ 	& ::= & 		$\term_{s} \in \transducerof{\term_{s}}$ \\
		%$\constr_{g}$ 	& ::= & 		$\term_{s} \in \cfg$ \\
		%$\constr_{i}$ 	& ::= & 		$\term_{i} \geq \term_{i}$ 	\\
		%$\term_{s}$		& ::= &  	$w$ 	$|$ $x$ 	$|$ 	$\term_{s} \app \term_{s}$														& \\
		%$\term_{i}$		& ::= &		$\lengthof{\term_s}	$ $|$ $k$	 \\					
		%\end{tabular}
$
		\begin{array}{c @{\,::=\ } l } 
		\formula 		& 		\constr \mid \formula \wedge \formula \mid \formula \vee \formula \mid \neg\formula\\
		\constr 			& 		\memconstrof \strterm \automaton \mid \relconstrof \strterm \strterm   \rel \mid \arterm \geq \arterm  \\ 
		\rel 	   		&	   \transducer \mid {=} \\
		\strterm			&     \emptystr \mid	x \mid \strterm \concat \strterm\\														
		\arterm 			&		k \mid \lengthof{\strterm} \mid \arterm + \arterm 	 					
		\end{array}
$
};
\end{tikzpicture} 
\vspace{-8mm}
\captionof{figure}{Syntax of string formulae}
\label{table:syntax}
\end{wrapfigure}

The syntax of a string formula $\formula$ over an alphabet $\alphabet$ and a set of variables $\varset$ 
is given in Figure \ref{table:syntax}. 
It is a Boolean combination of memberships, relational, 
and arithmetic constraints over string terms $\strterm$ (i.e., 
concatenations of variables in  $\varset$). 
\emph{Membership constraints} denote
 membership in the language of a finite-state automaton $\automaton$ over $\alphabet$. 
\emph{Relational constraints} denote either an equality of string terms, 
which we normally write as $\term = \term'$ instead of $\relconstrof \term {\term'} {{=}}$, 
or that the terms are related by a relation recognised by a transducer $\transducer$. Observe that the equality relations, that is, word equations, can be also expressed using length preserving transducers. 
In other words, the $=$ is only a syntactic sugar, a special case of $\transducer$.
Finally, arithmetic terms $\arterm$ are linear functions over term lengths and integers, 
and arithmetic constraints are inequalities of arithmetic terms.
String formulae allow using negation with one restriction, 
namely, constraints that are \emph{not invertible} (negation can be eliminated) must have only positive occurrences. 
It is well known that general transducers are not invertible, 
%it is not possible to negate them. 
%
while regular membership, length preserving relations (including equality), and length constraints are invertible.

To simplify presentation, 
we do not consider \emph{mixed} string terms $\strterm$ that contain, besides variables of $\varset$, also symbols of $\alphabet$. 
This is without loss of generality because a mixed term can be encoded as a conjunction of the pure term over $\varset$ obtained by replacing every occurrence of a letter $a\in\alphabet$ by a fresh variable $x$ and the regular membership constraints $\memconstrof x {\automaton_a}$ with $\langof{\automaton_a} = \{a\}$. Observe also that membership and equality constraints  may be expressed using transducers.

\medskip

\noindent
{\bf Semantics.}
We describe the semantics of our  logic using a mapping $\eta$, called \emph{interpretation}, that assigns to each string variable in $\varset$ a word in  $\alphabet^*$. Extended to string terms by 
%, $\eta(w) = w$ and 
$\eta(\term_{s_1} \concat \term_{s_2}) = \eta(\term_{s_1}) \concat \eta(\term_{s_2})$. Extended to arithmetic terms by $\eta(|\term_s|) = |\eta(\term_s)|$, $\eta(k) = k$ and $\eta(\term_i + \term_i') = \eta(\term_i) + \eta(\term_i')$. 
Extended to atomic  constraints, $\eta$ returns a truth value:
$$
\begin{array}{r @{\text{\ \ \ iff \ \ }} l}
\eta({\memconstrof \strterm  \automaton})=\top	 			 & \eta(\strterm)  \in \langof{\automaton} \\ 	
\eta({\relconstrof \strterm {\strterm'} \rel})=\top	 & (\eta(\strterm),\eta(\strterm')) \in \relof{\rel} \\ 	
\eta({\term_{i_1} \leq \term_{i_2}})=\top  				 & \eta{(\term_{i_1})}  \leq \eta{(\term_{i_2})}\nonumber 
\end{array}
$$
Given two interpretations $\eta_1$ and $\eta_2$ over two disjoint sets of string variables $\varset_1$ and $\varset_2$, respectively. We use $\eta_1 \cup \eta_2$ to denote the interpretation over $\varset_1 \cup \varset_2$ such that $(\eta_1 \cup \eta_2)(x)= \eta_1(x)$ if $x \in \varset_1$ and  $(\eta_1 \cup \eta_2)(x)= \eta_2(x)$ if $x \in \varset_2$.

The truth value of a Boolean combination of formulae under $\eta$ is defined as usual.
If $\eval{\formula} = \top$ then $\eta$ is a \emph{solution} of $\formula$,  
written $\eta \models \formula$. 
The formula $\formula$ is {\it satisfiable} iff it has a solution, otherwise it is {\it unsatisfiable}.

A relational constraint is said to be  {\em left-sided} if and only if it is on the   form $\relconstrof x \strterm   \rel $   where $x \in \varset$ is a string variable and $\strterm$ is a string term. Any string formula can be transformed into a formula where all the relational constraints are   left-sided by replacing any relational constraint of the form $\relconstrof \strterm {\strterm'}   \rel $ by ${\relconstrof x {\strterm'}   \rel } \wedge  x = {\term}$ where $x$ is fresh. 

A  formula $\formula$ is said to be \emph{concatenation-free} if and only if for every relational constraint $\relconstrof \strterm {\strterm'} \rel$, the string terms $\strterm$ and $\strterm'$ appearing in the  parameters of any relational constraints  in $\formula$ are   variables (i.e., $\strterm, \strterm' \in \varset$).

\section{Chain Free and Weakly Chaining Fragment}
It is well known that the satisfiability problem for the class of string constraint formulas is undecidable in general \cite{morvan,Chen:2019}. This problem is undecidable already for a single transducer constraint of the form ${\relconstrof x x \transducer}$ (by a simple reduction from the Post-Correspondence Problem). In the following, we define a subclass called {\em weakly chaining fragment} for which we prove that the satisfiability problem is decidable.

\paragraph {Splitting graph.}
Let $\formula \defeq \bigwedge_{j=1}^m\constr_j$ be a conjunction of relational string constraints with 
$\constr_j \defeq \relconstrof {t_{2j-1}} {t_{2j}} {\rel_j}, 1\leq j \leq m$ 
where for each $i:1\leq i \leq 2m$, 
$t_i$ is a concatenation of variables $x_i^1\concat\cdots\concat x_i^{n_i}$.
We define the set of \emph{positions} of $\formula$ as 
$P = \{\posof i j\ |\ 1 \leq j \leq 2m\ \land\ 1 \leq i \leq n_j\}$. 
The \emph{splitting graph} of $\formula$ is then the graph 
$G_\formula = (P, E, \nodevar, \edgecons)$
where the positions in $P$ are its nodes, 
and the mapping $\nodevar:P\rightarrow \varset$ labels each position $\posof i j$ with the variable $x^i_j$ appearing at that position. 
We say that $(i,2j-1)$  (resp. $(i,2j)$) is the $i$th \emph{left} (resp.  \emph{right}) positions of the $j$th constraint, and that $R_j$ is the predicate of these positions. 
Any pair of a left and a right position of the same constraint are called \emph{opposing}.
The set $E$ then consists of edges $(\pos,\pos')$ between positions for which there is an position $\pos''$ different from $\pos'$ that is opposing to $\pos$ and is labeled by the same variable as $\pos'$, i.e $\var(\pos'') = \var(\pos')$. 
Finally, the labelling $\edgecons$ of edges assigns to $(\pos,\pos')$ the constraint of $\pos$, 
that is, $\edgecons(\pos,\pos') = \rel_j$ where $\pos$ is a position of the $j$th constraint.
An example of a splitting graph is on Fig.~\ref{fig:graph}.
%
%
%\lukas{do we need the $\edgecons$?}

\paragraph{Chains.}
A \emph{chain}\footnote{We use chains instead of cycles or loops in order to avoid confusion between our decidable fragment and the ones that exist in literatures.} in the graph is a sequence $(\pos_0,\pos_1),(\pos_1,\pos_2),\ldots,(\pos_n,\pos_0)$ of edges in $E$.
A chain is \emph{benign} if $(1)$ all the relational constraints corresponding to the edges $\edgecons(\pos_0,\pos_1),\edgecons(\pos_1,\pos_2), \ldots,\edgecons(\pos_n,\pos_0) $  are left sided and all the string relations involved in these constraints are length preserving, and $(2)$ the sequence of positions $p_0,p_1, \ldots,p_n$ consists  of left positions only, or from right positions only. Observe that if there is a benign chain that uses only right positions then there exists also a  benign chain that uses only left positions.
The graph is \emph{chain-free} if it has no chains, and it is \emph{weakly chaining} if all its chains are benign. 
A formula is \emph{chain-free} (resp.  \emph{weakly chaining}) if the splitting graph of every clause in its DNF is chain-free (resp. weakly chaining).
Benign chains are on Fig.~\ref{fig:graph} shown in red.

In the following sections, we will show decision procedures for the chain-free and weakly chaining fragments.
Particularly, we will show how a weakly chaining formula can be transformed to a chain-free formula by elimination of benign cycles,
how then concatenation can be eliminated from a chain-free formula, and finally how to decide a concatenation-free formula.

\medskip

\noindent
{\bf Undecidability of Chaining Formulae.}
Before presenting the decision procedures for weakly chaining formulae, 
we finish the current section by  stating that the chain-free fragment is indeed the limit of decidability of general transducer constraints, in the following sense: 
We say that two conjunctive string formulae have the same  \emph{relation-concatenation skeleton} if one can be obtained from the other by removing membership and length constraints and replacing a  constraint of the form $\rel(t,t')$ by another constraint of the form $\rel'(t,t')$.
A \emph{skeleton class} is then an equivalence class of string formulae that have the same relation-concatenation skeleton.
Presence of chains depends only on the skeleton class, hence we call chaining or chain free also the skeleton classes of constraints of the respective property.

\begin{lemma}\label{lemma:undecidability}
%The   satisfiability problem is undecidable  for every  given \emph{skeleton class}.
\lukas{In the conference version the "of chaining constraints" was missing in the lemma statement, so it was intelligible. Add footnote about this?}
The satisfiability problem is undecidable for every fixed skeleton class of chaining constraints.
\end{lemma}
Together with decidability of chain-free formulae, discussed in Sections \ref{sec:splitting} and \ref{sec:concat-free}, the lemma implies that  the satisfiability problem for a  skeleton class is decidable \emph{if and only if} its splitting graph is chain-free.
In other words, chain-freeness is the most precise criterion of decidability of string formulae based on relation-concatenation skeletons
(that is, a criterion independent of the particular values of relational, membership, and length constraints).

%The proof of the above lemma can be done through  a reduction from undecidability of general transducer constraints of the form $\transducer(x,x)$.
%
%\lukas{included from appendix}
%%%%%%%%%%%%%% BEGIN JOURNAL UNDECIDABILITY PROOF %%%%%%%%%%%%%%%%%%%%%%%%%%%%%%%%%%%%%%
\begin{proof}[Lemma~\ref{lemma:undecidability}]
\newcommand{\id}{\mathit{Id}}
We will show the lemma by reduction of undecidability of constraints of the form $\transducer(x,x)$  %\cite{morvan,Chen:2019}\lukas{is the reference right?}.
Let $\Phi$  be a chaining conjunction of relational constraints.
We will show how any constraint $\transducer(x,x)$ can be translated into an equivalent constraint $\Phi'$ with the same skeleton as $\Phi$.

Let $\Psi = \rel_1(\term_1,\term_1') \land \cdots \land \rel_n(\term_n,\term_n')$ be a conjunction of a smallest sub-set of the relational constraints of $\Phi$ which is chaining.
Let $x_i, 1\leq i \leq n$, be the variable which labels the position within $t_i$ that is a part of the chain. 
$\Phi'$ will be constructed to be equivalent to 
$\transducer(x_1,x_{2}) \land x_{2} = x_3 \land \cdots \land x_n = x_1$.

Let $\sharp$ be a symbol outside the alphabet $\Sigma$, let $\Sigma_\sharp$ denote $\Sigma\cup\{\sharp\}$. 
%
%Let $k_i$ be the number of occurrences of $x_i$ in $t_i$ and let $l_i$ be the number of occurrences of $x_i$ in $t_{i-1}'$ or in $t_n'$ if $i = 1$.
%
Then the new constraint $\Phi'$ is created from $\Psi$ by:

\begin{enumerate}
%\item 
%introducing the membership constraint $x_i\in \automaton_{\Sigma^*.\{\sharp\}}$ with $\langof{\automaton_{\Sigma^*.\{\sharp\}}} = \Sigma^*.\{\sharp\}$ for all $1\leq i \leq n$,
%\item 
%introducing a membership constraint $y\in \automaton_\epsilon$ with $\langof{\automaton_\epsilon} = \{\epsilon\}$ for all variables $y$ other then $x_1,\ldots,x_n$.
\item 
Introducing membership constraints $x_i\in \automaton, 1\leq i \leq n$ with $\langof{\automaton} = \Sigma^*\concat\{\sharp\}$ (this is to make the borders of the variables $x_i$ visible).
\item 
Introducing membership constraints $y\in \automaton_\epsilon$ with $\langof{\automaton_\epsilon} = \{\epsilon\}$ for all variables $y$ other then $x_1,\ldots,x_n$ (this is to ``erase'' these variables from the valuations of $t_1,\ldots,t_n$).
\item 
Replacing every $\rel_i$ in $\Psi$ by $\rel_i'$ where  
\begin{enumerate}
\item 
$\relof{\rel_1'} = \relof{\transducer} \concat \{(\sharp,\sharp)\} \concat \Sigma_\sharp^*\times \Sigma^*_\sharp$ and
\item
%$\relof{\rel_i'} = \id \concat \{(\sharp,\sharp)\} \concat \Sigma_\sharp^*\times \Sigma_\sharp^*$ for $1 < i \leq n$, $\id$ being the identity on $\Sigma^*$. 
$\relof{\rel_i'} = \id \concat \{(\sharp,\sharp)\} \concat \Sigma_\sharp^*\times \Sigma_\sharp^*$ for $1 < i \leq n$, $\id$ is the identity on $\Sigma^*$. 
\end{enumerate}
Here (a) together with (1) and (2) assures that the first track of $\transducer$ is mapped at the first occurence of $x_1$ in $t_1$ and the second track at the first occurence of $x_2$ in $t_2$ (it cannot be multiple occurrences of $x_1$ or of $x_2$ due to the use $\sharp$ as a variable delimiter in (1) and (3a)); and (3b) together with (1) and (2) encodes the equalities $x_i = x_{i+1}$, since they ensure that the left track of the identity transducer in $\rel_i'$ is mapped at $x_i$ and the right track at $x_{i+1}$. 
\item 
Replacing all the relational constraints $\rel(t,t')$ of $\Phi$ outside $\Psi$ by $\rel'(t,t')$ where $\relof{\rel'} =\Sigma^*\times\Sigma^*$ (to make them irrelevant).
\end{enumerate}
%
%The constraint $\Phi'$ was designed for this to hold, particularly,
%the role of the $\automaton_\epsilon$-membership constraints was to eliminate assignments to variables other then $x_1,\ldots,x_n$,
%the part of the primed relational constraints $\{(\sharp,\sharp)\} \concat \Sigma_\sharp^*\times \Sigma^*_\sharp$ was added to eliminate the effect of possible multiple occurrences of the variables $x_i$ in terms,
%and the $\Sigma^*\times\Sigma^*$ constraints outside $\Psi$ were chosen to eliminate any influence of the part of $\Phi$ outside the chain.

Last, to show that 
$\Phi' \equiv \transducer(x_1,x_{2}) \land x_{2} = x_3 \land \cdots \land x_n = x_1$ indeed holds,
it must be argued that every $x_i,2\leq i \leq n$ appears in $\Psi$ only in $t_i$ and $t'_{i-1}$ and
$x_1$ appears in $\Psi$ only in $t_1$ and $t'_{n}$ (otherwise the cycle of variable occurences related by $\transducer$ and $\id$ by relations from point (3) above could be cut short, resulting in a shorter cycle of occurences related by $\id$ only).
This may be shown from the minimality of $\Psi$, by contradiction: if $x_i$ appeared somewhere else, we could shorten the chain, and $\Psi$ would not be minimal. Particularly, take w.l.o.g. $x_1$ as the variable contradicting the claim. There are two options:
\begin{itemize}
\item
$x_1$ appears also in some $t_m, 1< m \geq n$. Then, in the case $m < n$, the splitting graph has an edge from $x_1$ in $t_m$ to the $x_{m+1}$ in $t_{m+1}$ and $x_n$ in $t_n$ has an edge to the $x_1$ in $t_m$, closing a shorter chain that bypasses the first $m-1$ edges of the original. Otherwise, if $m=n$, then due to the occurence of $x_1$ in $t_{n'}$, there is a self-loop on the occurence of $x_1$ in $t_n$ (as well as on the occurence in $t_{n'}$).
Both cases contradict the assumed minimality of $\Psi$.
\item
$x_1$ appears also in some $t_m',1 \leq m > n$. But then the splitting graph has an edge from the occurrence of $x_m$ in $t_m$ to the occurrence of $x_1$ in $t_1$, closing a shorter chain that bypasses the last $n-m +1$ edges of the original, which again contradicts the minimality of $\Psi$.
\end{itemize}%
\end{proof}

%%%%%%%%%%%%%  END JOURNAL UNDECIDABILITY PROOF %%%%%%%%%%%%%%%%%%%%%%%%%%%%%%%%%%%%%%%%
\section{Weakly Chaining to Chain-Free}
In the following, we show that, given a weakly chaining formula, we can transform it to an equisatisfiable  chain-free formula. 

\begin{theorem}
\label{thm-chainfree}
A weakly chaining formula can be transformed to an equisatisfiable chain-free formula.
\end{theorem}

The rest of this section is devoted to the proof of Theorem \ref{thm-chainfree} (which also provides to transform any weakly chaining formula  into an equisatisfiable chain-free formula). In the following, we  assume w.l.o.g. that the given weakly-chaining formula $\formula $ is conjunctive. The proof is done by induction on the number ${\bf B}$ of relational constraints that are labelling the set of benign chains in the splitting graph of ${\formula}$.

\smallskip

\noindent
{\em Base Case ({\bf B=0}).} Thus there is no benign chain in $G_{\formula}$ and so  Theorem \ref{thm-chainfree}  holds.

\smallskip

\noindent
{\em Induction Case ({$\bf B>0$}).}
Let us start by showing how to remove one benign chain (and the set of relational constraints labelling it) in the case where the splitting graph of $\formula$ does not contain nested chains. Let $\rho = (\pos_0,\pos_1),(\pos_1,\pos_2),\ldots,(\pos_n,\pos_0)$ be a benign chain in the splitting graph $G_\formula$. For every $i \in \{0,\ldots,n\}$, let $\relconstrof {x_i} {t_{i}} {\rel_i}$ be the length preserving relation constraint to which the position $\pos_i$ belongs. We assume w.l.o.g.\footnote{This is possible since   if there is benign chain that uses only right positions then there exists also a  benign chain that uses only left positions.} that all the positions $\pos_0, \pos_1,\ldots, \pos_n$ are left positions. Since $\rho$ is a benign chain, the variable $x_i$ is appearing in the string term $t_{(i-1) \modulo (n+1)}$%
\footnote{We write $x \mod y$ for the number in $\{0,\ldots,y-1\}$ equivalent to $x$ modulo $y$. Note that it is always a positive number, even when $x$ is negative.}
for all  $i \in \{0,\ldots,n\}$. Furthermore, we can use the fact that the relational constraints are length preserving to deduce that the string-values of variables $x_0,x_1, \ldots, x_n$ must have the same length (it is a crucial insight for the proof). This  implies also that, for every $i \in \{0,1,\ldots,n\}$, the string term $t'_i$, that is constructed by removing from $t_i$ one occurrence of the variable $x_{(i+1) \modulo (n+1)}$, is equivalent to the empty word. Therefore,  the relational constraint  $\relconstrof {x_i} {t_{i}} {\rel_i}$ can be rewritten as 
 $\relconstrof {x_i} {x_{(i+1) \modulo (n+1)}} {\rel_i}$ for all $i \in \{0,1,\ldots,n\}$.
 
Let $x_{i_1},x_{i_2},\ldots, x_{i_k}$ be the maximal subsequence of pairwise distinct variables in
$x_0,x_1,\ldots,x_n$. Let ${\tt index}$ be a  mapping that associates to each index $\ell \in
\{0,\ldots,n\}$, the index $j \in \{1,\ldots,k\}$  such that $x_\ell=x_{i_j}$. We can  transform the
transducer $\rel_i$, with $i \in \{0,\ldots,n\}$, to a length preserving $k$-tape automaton $A_i$
such that a word $(w_1,w_2,\ldots,w_k) $ is accepted by $A_i$ if and only if $(w_{{\tt
index}(i)},w_{{\tt index}((i+1) \modulo (n+1))})$ is accepted by $\rel_i$. Then, let $A$ be the $k$-tape
automaton resulting from the intersection of $A_0,\ldots, A_n$. Observe that $A$ is also a
length-preserving automaton. Furthermore, we have that the tupe of words $(w_1,w_2,\ldots,w_k) $ is accepted by $A$ if
and only if 
for all $i \in \{0,\ldots,n\}$,
$(w_{{\tt index}(i)},w_{{\tt index}((i+1) \modulo (n+1))})$ is accepted by $\rel_i$
(i.e., the automaton $A$ characterizes all possible solutions of
$\bigwedge_{i=0}^n \relconstrof {x_i} {x_{(i+1) \modulo (n+1)}} {\rel_i}$). Ideally, we would like to
replace the $\bigwedge_{i=0}^n \relconstrof {x_i} {x_{(i+n) \modulo (n+1)}} {\rel_i}$ by
$A(x_{i_1},x_{i_2},\ldots, x_{i_k})$, however, our syntax forbids $k$-ary relations.  To circumvent this, we extend our alphabet $\Sigma$ by all the letters in $\Sigma^k$ and
replace  $\bigwedge_{i=0 }^n\relconstrof {x_i} {x_{(i+1) \modulo (n+1)}} {\rel_i}$ by  $\varphi:= A(x)
\wedge \bigwedge_{j=1 }^k\relconstrof {x_{i_j}} {x} {\pi_j}$ with a fresh variable $x$ and for
every $j \in \{1,\ldots,k\}$, $\pi_j$ the length preserving transducer accepting all pairs of the form $(w_j,(w_1,w_2,\ldots,w_k))$. 
Finally, let $\formula'$ be the formula that is obtained from
$\formula$ by replacing the subformula $\bigwedge_{i=0}^n \relconstrof {x_i} {t_{i}} {\rel_i}$ in
$\formula$ by $\varphi \wedge |t'_i|=0$ (remember that the string term $t'_i$ is  $t_i$ with one
occurrence of the variable  $x_{(i+1) \modulo (n+1)}$ removed).  It can be seen from the discussion above
that $\bigwedge_{i=0}^n \relconstrof {x_i} {t_{i}} {\rel_i}$ is equivalent to $\exists x:\varphi
\wedge |t'_i|=0$ (both say that $\relconstrof {x_i} {x_{(i+1) \modulo (n+1)}} {\rel_{i}}$ holds for all
$i\in\{0,\ldots, n\}$ and that values of all the other variables occurring in $t_0,\ldots,t_n$ must be
$\epsilon$). Hence $\formula'$ and $\formula$ are equisatisfiable. Furthermore, the number of
relational constraints that are labelling the set of benign chains in the \mbox{splitting graph} of
$\formula'$ is strictly less than ${\bf B}$ (since $\pi_j$ cannot label any \mbox{benign chain in
$\formula'$).}

Let us now consider the case where the splitting graph of $\formula$ contains nested benign chains.
Let $\rho = (\pos_0,\pos_1),(\pos_1,\pos_2),\ldots,(\pos_n,\pos_0)$ be a benign chain in the
splitting graph $G_\formula$. We assume w.l.o.g. that all the positions $\pos_0, \pos_1,\ldots,
\pos_n$ are left positions. Let $S$ be \emph{the smallest} set of positions in $G_\formula$
satisfying that: (1) $\{\pos_0,\pos_1,\ldots,\pos_n\}$ is a subset of $S$, and (2) $S$ includes all
positions of every benign chain that traverses a position $\pos \in S$. Assume that $S =
\{q_0,q_1,\ldots,q_m\}$, and  observe that all the positions in $S$ are left positions and for each
pair of positions $q_i$ and $q_j$, there is a path in the graph  $G_\formula$ from $q_i$ to $q_j$.
For every $i \in \{0,\ldots,m\}$, let $\relconstrof {x_i} {t_{i}} {\rel_i}$ be the length preserving
relation constraint to which the position $q_i$ belongs. Using a similar reasoning as in the case of
non-nested chains, we can deduce that all the string-values of variables $x_i$ must have the same
length. Let $y_i$ be the {\em unique} string variable in $\{x_0,x_1,\ldots,x_m\}$ that appears in
$t_i$ and let the rest of $t_i$, i.e. $t_i$ with $y_i$ removed, be denoted $t_i'$ (The uniqueness of $y_i$ can be assumed without loss of generality. Observe that if there were
multiple occurrences of variables from $\{x_0,x_1,\ldots,x_m\}$ in $t_i$, then, since their values
must have the same length as that of $x_i$, the only length possible was $0$, i.e. all variables
$x_0,x_1, \ldots,x_m$ must be assigned $\epsilon$. We could replace them in $\formula$ by $\epsilon$
and remove $\bigwedge_{i \in \{0,\ldots,m\}} \relconstrof {x_i} {t_{i}} {\rel_i}$ from $\formula$.).  
%Let  $t'_i$ be the string term constructed from $t_i$ by removing all variables but $y_i$. 
Notice that all variables of each $t_i$ other then $y_i$ must be valuated by $\epsilon$, because, again, the relations are length preserving and the constraints are left-sided. 
%from $t_i$ the variable $y_i$, is equivalent to the empty word. 
Therefore, these variables can be removed from the terms and each relational constraint  $\relconstrof {x_i} {t_{i}} {\rel_i}$, $i \in \{0,1,\ldots,m\}$, can be rewritten as 
$\relconstrof {x_i} {y_i} {\rel_i}$.

Next, we can proceed in similar manner as in the case of non-nested chains. Let $x_{i_1},x_{i_2},\ldots, x_{i_k}$ be the maximal subsequence of pairwise distinct variables taken from $x_0,x_1,\ldots,x_m$. Let ${\tt index}_x$ (resp. ${\tt index}_y$) be a  mapping that associates to each index $\ell \in \{0,\ldots,m\}$, the index $j \in \{1,\ldots,k\}$  such that $x_\ell=x_{i_j}$ (resp. $y_\ell=x_{i_j}$). We can  transform the transducer $\rel_i$, with $i \in \{0,\ldots,m\}$, to a length preserving $k$-tape automaton $A_i$ such that a word $(w_1,w_2,\ldots,w_k) $ is accepted by $A_i$ if and only if $(w_{{\tt index}_x(i)},w_{{\tt index}_y(i)})$ is accepted by $\rel_i$. Let $A$ be the length preserving $k$-tape automaton resulting from the intersection of $A_0,\ldots, A_m$. Then, we have that $(w_1,w_2,\ldots,w_k) $ is accepted by $A$ if and only if $(w_{{\tt index}_x(i)},w_{{\tt index}_y(i)})$ is accepted by $\rel_i$ for all $i \in \{0,\ldots,m\}$ (i.e., the automaton $A$ characterizes all possible solutions of $\bigwedge_{i=0 }^m\relconstrof {x_i} {y_i} {\rel_i}$). 
Now, we are ready to transform the subformula $\bigwedge_{i=0} ^m\relconstrof {x_i} {y_{i}} {\rel_i}$ in $\formula$ into a chain-free formula. To do that,  we first  extend our alphabet $\Sigma$ by all the letters in $\Sigma^k$ and then we replace  $\bigwedge_{i= 0}^m\relconstrof {x_i} {y_i} {\rel_i}$ by  $\varphi:= A(x) \wedge \bigwedge_{j\in \{1,\ldots,k\} }\relconstrof {x_{i_j}} {x} {\pi_j}$ where $x$ is a fresh variable and  for every $j \in \{1,\ldots,k\}$, the transducer $\pi_j$ is  defined as in the non-nested case. Finally, let $\formula'$ be the formula obtained from $\formula$ by replacing the sub-formula $\bigwedge_{i \in \{0,\ldots,n\}} \relconstrof {x_i} {t_{i}} {\rel_i}$ in $\formula$ by $\varphi \wedge |t'_i|=0$ (remember that is the string term $t'_i$ is  $t_i$ from which we have removed one variable  from $\{x_0,x_1,\ldots,x_m\}$). Now we are ready to state our lemma:

\begin{lemma}
\label{removingchains}
The formula $\formula$ is satisfiable if and only if the formula   $\formula'$ is satisfiable. Furthermore, if $\sigma:=(s_0,s_1),(s_1,s_2),\ldots,(s_\ell,s_0)$ is a chain in  the splitting graph of $\formula'$ then $\sigma$ is also a chain in the splitting graph of $\formula$.\end{lemma}

%\section{Proof of Lemma~\ref{removingchains}}
\begin{proof}
The proof of the equisatisfiability between $\formula$ and $\formula'$ is an immediate consequence
of the construction. Let us prove by contradiction the second part of the lemma. Let us assume that
there is a chain $\sigma:=(s_0,s_1),(s_1,s_2),\ldots,(s_\ell,s_0)$ in the splitting graph of
$\formula'$ that is not a chain in the splitting graph of $\formula$. This means that at least one
of the edges of the chain $\sigma$ is labelled by a relational constraint in $\varphi$ (i.e.,
$\{\pi_1, \pi_2,\ldots,\pi_k\} \cap \{\edgecons(s_0,s_1),\edgecons(s_1,s_2),
\ldots,\edgecons(s_\ell,s_0) \}\neq \emptyset$). On the other hand, there must be edges labeled by constraints other than $\pi_o,1\leq o \leq k$ since $G_\varphi$ cannot have a chain, by definition.
%
%Let $j \in \{0,1,\ldots,\ell\}$  (resp. $j' \in
%\{0,1,\ldots,\ell\}$) be the smallest (resp. largest) index such that $\edgecons(s_j,s_{(j+1) \modulo
%(\ell+1)})$ (resp. $\edgecons(s_{j'},s_{({j'}+1) \modulo (\ell+1)})$) is in  $\{\pi_1,
%\pi_2,\ldots,\pi_k\}$. 
%
%Let $j \in \{0,1,\ldots,\ell\}$ be the smallest index such that $\edgecons(s_j,s_{(j+1) \modulo
%(\ell+1)})\in\{\pi_1,
%\pi_2,\ldots,\pi_k\}$ and 
%let $j' \in
%\{0,1,\ldots,\ell\}$ be the largest such index. 
%
Let $j,j'\in \{0,\ldots,\ell+1\}$ mark the borders of a maximum sub-sequence of the chain $(s_{j},s_{(j+1) \modulo (\ell+1)}),$ $(s_{(j+1) \modulo (\ell+1)},s_{(j+2) \modulo (\ell+1)}), \ldots,(s_{(j'-1) \modulo (\ell+1)},s_{j'})$ in which constraints on all edges are in $\{\pi_1, \pi_2,\ldots,\pi_k\}$.
%\footnote{Note that $j+\ell \modulo (\ell +1)$ is equivalent to $(j-1) \modulo (\ell+1)$.}.
%
%Let $j \in \{0,1,\ldots,\ell\}$ be an index such that 
%$\edgecons(s_j,s_{(j+1) \modulo (\ell+1)})\in\{\pi_1, \pi_2,\ldots,\pi_k\}$ and 
%$\edgecons(s_{(j+\ell) \modulo (\ell+1)},s_{j})\not\in\{\pi_1,\pi_2,\ldots,\pi_k\}$.\footnote{Note that $j+\ell \modulo (\ell +1)$ is equivalent to $(j-1) \modulo (\ell+1)$.}
%Let $j' \in
%\{0,1,\ldots,\ell\}$ be the index such that 
%$\edgecons(s_{j'+1 \modulo (\ell+1)}\not\in\{\pi_1,\pi_2,\ldots,\pi_k\})$ and has 
%the minimum distance from $j$, where the distance from $j$ is the number $d\in\{0,\ldots,\ell\}$ such that $j' = (j+d) \modulo (\ell +1)$ 
% (that is, $s_j$ and $s_{j'}$ are the borders of a maximum interval of the chain, modulo $\ell+1$, where all constraints are in $\{\pi_1, \pi_2,\ldots,\pi_k\}$). 
%
%Let us assume that $\pi_i:= \edgecons(s_j,s_{(j+1) \modulo (\ell+1)})$ (resp.
%$\pi_{i'}:= \edgecons(s_{j'},s_{({j'}+1) \modulo (\ell+1)})$).  
%
Let us assume that $\pi_i:= \edgecons(s_j,s_{(j+1) \modulo (\ell+1)})$ and 
$\pi_{i'}:= \edgecons(s_{j'},s_{({j'}+1) \modulo (\ell+1)})$.  
%
%It is easy to see that $\nodevar(s_j)=x_i$  (resp. $\nodevar(s_{j'})=x$). 
%
Than $\nodevar(s_j)=x_i$ and $\nodevar(s_{j'})=x$ (since positions with $x$ appear only in constraints $\pi_o(x_o,x)$, i.e., edges from or to $x$ may only be labeled by some $\pi_o$, and these cannot lead to $s_j$ or from $s_{j'}$ by definition of $j$ and $j'$). 
This means that the  position $s_j$ is a left
position and corresponds to the variable $x_i$ while the position $s_{j'}$ is a right position and
corresponds to the fresh variable $x$. Observe that $\nodevar({s_{(j'+1) \modulo (\ell+1)}})=x_{i'}$.
Furthermore, we know that the position $s_{(j-1) \modulo (\ell+1)}$ belongs to the splitting graph
$G_{\formula}$ (as well as to $G_{\formula'}$). This implies that $s_{(j-1) \modulo (\ell+1)}$ is not in the set $S$ (otherwise it will
not belong to the graph $G_{\formula'}$). By definition, the subgraf of $G_\formula$ induced by $S$ is strongly connected. Therefore, there is
a path $(q_{j_0},q_{j_1}),(q_{j_1},q_{j_2}),\ldots,(q_{j_{r-1}},q_{j_r})$ in $G_\formula$, that uses only positions in $S$ s.t.  $\nodevar{(q_{j_0})}=x_i$ and
$\nodevar{(q_{j_r})}= x_{i'}$. Then, by
replacing the path\\ $(s_{(j-1) \modulo (\ell+1)},s_j)(s_j,s_{(j+1) \modulo (\ell+1)}),\ldots ,(s_{j'
},s_{(j'+1) \modulo (\ell+1)})$  in  $\sigma$  by\\ $(s_{(j-1) \modulo
(\ell+1)},q_{j_0}),(q_{j_0},q_{j_1}),(q_{j_1},q_{j_2}),\ldots,  (q_{j_{r-1}},s_{(j'+1) \modulo
(\ell+1)})$, we obtain another chain $\sigma'$. Since $\sigma'$ is a chain in
$G_{\formula}$, it is a benign chain. Since $\sigma'$ traverses some
positions in $S$, all the positions of $\sigma'$ are in $S$
including the position $s_{(j-1) \modulo (\ell+1)}$, which is a contradiction.
\end{proof}

As a corollary of the above lemma, we obtain that the new relation $\pi_j$ cannot be part of any benign chain. Thus, the number of relational constraints that are labelling the set of benign chains in the \mbox{splitting graph} of $\formula'$ is strictly smaller than  ${\bf B}$ (since $\pi_j$ cannot be used to label any benign chain in $\formula'$).

%%%%%%%%%%%%%% END JOURNAL BENIGN TO CHAIN-FREE PROOF %%%%%%%%%%%%%%%%%%%%%%%%%%%%%%%%%%%%%%

\section{Chain-Free to Concatenation Free} 
\label{sec:splitting}

In the following, we show that we can reduce the satisfiability problem for a chain free formula to the satisfiability problem of a concatenation-free formula. To that end, we describe an algorithm that eliminates concatenation from relational constraints by iterating simple splitting steps. 
When it terminates, it returns a formula over constraints that are concatenation-free.
The algorithm can be applied if the string constraints in the formula \emph{allow splitting}, 
and it is guaranteed to terminate if the formula is \emph{chain-free}. 
We will explain these two conditions below together with the description of the algorithm.

\medskip

%\noindent
%{\bf Splitting.} 
\subsection{Split of a Constraint}
The \emph{split} of a relational constraint $\constr \defeq \relconstrof {x\concat\term} {y\concat\term'} \rel$ with $\term,\term'\neq\epsilon$ 
is the formula 
$\leftsplit \lor \rightsplit$ where
\begin{eqnarray}
\label{eq:lsplit}
\leftsplit  \defeq & \bigvee_{i=1}^n \relconstrof {x_1} {y} {R_i}\land \relconstrof {x_2\concat\term} {\term'} {R_i'} \ [x/x_1\concat x_2]\\
\label{eq:rsplit}
\rightsplit  \defeq & \bigvee_{j=1}^m \relconstrof x {y_1}{R_j}\land \relconstrof {\term}{y_2\concat\term'} {R_j'} \ [y/y_1\concat y_2]
\end{eqnarray}
$m, n\in\nat$,  $x_1,x_2,y_1,y_2$ are fresh variables, and  $\eta\models \constr$ 
if and only if there is an assignment $\eta':\{x_1,x_2,y_1,y_2\}\rightarrow \alphabet^*$ such that 
$\eta\cup\eta'\models (\leftsplit\land x = x_1\concat x_2) \lor (\rightsplit \land y = y_1\concat y_2)$.
The formula $\leftsplit$ is called the \emph{left split} and $\rightsplit$ is called the \emph{right split} of $\constr$.
In case $\term'=\epsilon$, the split is defined in the same way but with $\leftsplit$ left out,
and if $\term = \epsilon$, then $\rightsplit$ is left out. 
Namely, if $\constr \defeq \relconstrof {x\concat\term} {y} \rel$ then the split of $\clause$ is only
$$ 
%\begin{eqnarray}
%\label{eq:lsplit}
%\leftsplit  \defeq & \bigvee\nolimits_{i=1}^n \relconstrof {x_1} {y} {R_i}\land \relconstrof {x_2\concat\term} {\term'} {R_i'} \ [x/x_1\concat x_2]\\
%\label{eq:rsplit}
%\leftsplit \defeq \mathit{false} \quad\quad 
\rightsplit  \defeq  \bigvee_{j=1}^m \relconstrof x {y_1}{R_j}\land \relconstrof {\term}{y_2} {R_j'} \ [y/y_1\concat y_2] 
%\end{eqnarray}
$$
and if $\constr \defeq \relconstrof {x} {y\concat\term'} \rel$ then the split of $\clause$ is only
$$
%\begin{eqnarray}
%\label{eq:lsplit}
\leftsplit  \defeq \bigvee_{i=1}^n \relconstrof {x_1} {y} {R_i}\land \relconstrof {x_2} {\term'} {R_i'} \ [x/x_1\concat x_2]  
%\quad\quad \rightsplit  \defeq  \mathit{false} 
%\label{eq:rsplit}
%\rightsplit  \defeq & \bigvee\nolimits_{j=1}^m \relconstrof x {y_1}{R_j}\land \relconstrof {\term}{y_2} {R_j'} \ [y/y_1\concat y_2]
%\end{eqnarray}
%
$$
If both $\term$ and $\term'$ equal $\epsilon$, then $\constr$ is concatenation-free and does not have a split.
A simple example is the equation $x y = z z$ with the split $(x_1 = z \land x_2 y = z)\lor (x = z_1 \land y = z_2 z_1 z_2)$.

A class of relational constraints $\mathcal C$ \emph{allows splitting} if for every constraint in $\mathcal C$ that is not concatenation-free, 
it is possible to compute a split that \mbox{belongs to  $\mathcal C$.} 
Equalities as well as transducer constraints allow splitting. 
A left split of an equality $x\concat \term = y \concat \term'$ is $x_1 = y \land x_2\concat \term = \term'$. 
A left split of a transducer constraint $\relconstrof{x\concat\term} {y\concat\term'}{\transducer}$ is the formula 
$$
\bigvee_{q\in Q} \relconstrof {x_1}{y}{\transducer_q} \land \relconstrof {x_2 \concat\term} {\term'} {{}_q\!\transducer}
$$ 
where $Q$ is the set of states of $\transducer$, and ${}_q\!\transducer$ and $\transducer_q$ are the $\transducer$ with the original set of initial and final states, respectively, replaced by $\{q\}$ (this is the automata splitting technique of \cite{string14} extended to transducers in \cite{BL16}). 
The right splits are analogous. 

\medskip

%\noindent
%{\bf Splitting algorithm.}
\subsection{Splitting Algorithm}
The \emph{splitting algorithm} for eliminating concatenation iterates \emph{splitting steps} on a formula in DNF. 
A \emph{splitting step} can be applied to a clause that can be written in the form $\clause \defeq \constr\land \formula$ where $\constr\defeq \rel(x\concat\term,y\concat\term')$. 
The splitting  replaces the clause by a DNF of the disjunction
$$
(\leftsplit\land\formula[x/x_1\concat x_2]\land |x| = |x_1| + |x_2|) \lor (\rightsplit \land\formula[y/y_1\concat y_2] \land |y| = |y_1| + |y_2|)
$$ 
where $\constr$ is the \emph{split constraint} and $\leftsplit$ and $\rightsplit$ are its left and right split, respectively.
The left or the right disjunct is omitted if $\term' = \emptystr$ or $\term=\emptystr$, respectively. 
The splitting step is not applied when both $\term$ and $\term'$ equal $\emptystr$, i.e. $\constr$ is concatenation-free. 
Notice that the DNF produced by the splitting step has clauses of the form 
$(D_L\land\formula[x/x_1\concat x_2]\land |x| = |x_1| + |x_2|)$ or 
$(D_R\land\formula[x/x_1\concat x_2]\land |x| = |x_1| + |x_2|)$ where $D_L$ resp. $D_R$ is a disjunct from the left resp. the right split of $\constr$ (c.f. Equation~(\ref{eq:lsplit}) and (\ref{eq:rsplit})). 

%For every relational constraint $\psi'$ of $\Omega$ 
%
%we define its \emph{predecessor} constraint $\psi$ in $\clause$ and call $\psi'$ the \emph{successor} of $\psi$. 
%
We will further call constraints 
$\constr[x/x_1\circ x_2]$ in $\Psi[x/x_1\circ x_2]$ and $\constr[y/y_1\concat y_2]$ in $\Psi[y/y_1\concat y_2]$
\emph{successors} of the constraint $\constr$ in $\Psi$,  
%$\psi[x/x_1\circ x_2]$ in  $\Psi[x/x_1\circ x_2]$ and of $\psi[y/y_1\concat y_2]$ in $\Psi[y/y_1\concat y_2]$ 
and also call every constraint in $\leftsplit$ and $\rightsplit$ the successor of $\varphi$. 
We also say that the right hand side of the successor of a constraint is a successor of the right hand side of the constraint and likewise for their left-hand sides. 
By \emph{descendant} we refer to the transitive closures of the relation of being a successor. 
In the context of a splitting graphs, we refer to the leftmost position of the left-hand side of $\varphi$ (that of $x$) as the \emph{left split position} and to the leftmost position of the right-hand side of $\varphi$  (that of $y$) as the \emph{right split position}.

In order to ensure termination, the algorithm applies splitting steps under the following regimen consisting of two phases:
\begin{description}
\item
[Phase 1] 
Phase 1  maintains each clause $\clause$ of a DNF of the string formula annotated with a \emph{reminder}, 
a subset $\reminder$ of $\clause$.\footnote{We sometimes treat clauses as \emph{sets} of constraints of which they are conjunctions.}
A splitting step can be then applied to a constrain $\constr$ of $\clause$ only if it is a \emph{root constraint} of $\reminder$, 
that is, it is a constraint of $\reminder$ and all positions at one of its sides are \emph{root nodes} of the splitting graph $G_\reminder$ of the reminder (nodes without incoming edges). 
%
%The reminder graphs are assigned to clauses as follows.
%%
%Initially, $H_\clause \defeq G_\clause$ for each clause $\clause$.
%After taking a splitting step, 
%the reminder graph of each new clause $\clause'$ is a sub-graph $H_{\clause'}$ of its splitting graph $G_{\clause'}$. 
%%
%Particularly, $H_{\clause'}$ contains only those constraints of $\clause'$ (their positions that is) 
%that are not concatenation-free and are successors of the constraints of $\clause$ that appear in $H_\clause$. 
%The newly created concatenation-free constraints do not propagate to $H_{\clause'}$. 

%The reminder is assigned to clauses as follows.
%
Initially, the reminder of a clause is obtained from the clause itself by removing all concatenation-free root constraints.
%Initially, the reminder of a clause is obtained from it by removing all concatenation-free constraints. 
After taking a splitting step from a clause $\clause$, 
the reminder $\reminder'$ of each resulting clause $\clause'$ by
1) removing constraints that are not successors of the reminder $\reminder$ of $\clause$ and 
%2) removing concatenation-free successors of the split constraint. 
2) removing concatenation-free root-constraints. 
%Note that step (2) removes 
%a) concatenation-free successors of the split constraint. 
%constraints that are of those constraints of $\clause'$  
%that are not concatenation-free successors of root constraints.
%
%concatenation-free and are successors of the constraints of $\clause$ that appear in $H_\clause$. 
%The newly created concatenation-free constraints do not propagate to $H_{\clause'}$. 

Phase 1 terminates when the reminders of all clauses are concatenation-free (possibly empty). 
%Phase 1 terminates when the reminders of all clauses are empty. 

\item
[Phase 2] Phase 2 then performs splitting steps in any order until all constraints are concatenation-free.
\end{description}

\begin{theorem}
\label{theorem:splitting}
When run on a chain-free formula, 
the splitting algorithm terminates with an equisatisfiable chain and concatenation-free formula.
\end{theorem}

Before presenting the full proof of Theorem~\ref{theorem:splitting} in Section~\ref{section:termination}, we provide a brief overview of the proof and the role of chain-freeness in it.

The main difficulty with proving termination of the splitting algorithm comes from the substitution of variables involved in the left and right split. 
In the case of left split, a step towards concatenation-freeness is made by removing one concatenation operator $\concat$ from the clause, 
since the terms $x\concat \term$ and $y\concat \term'$ are replaced by $x_1$, $y$, $\term'$, and $x_2\concat \term$.
However, the substitution of $x$ by $x_1\concat x_2$ in the reminder of the clause introduces as many new concatenations as there were occurrences of $x$ other than the one explicit in the definition of the left split. 
The case of the right split is analogous.
There is hence no immediate guarantee that a sequence of left splits converges to a concatenation-free formula.
Termination is achieved by limiting the effects of the substitution through the assumption of chain freeness and the use of the splitting regimen defined by the splitting algorithm.

The role of chain-freeness in this is as follows.
The splitting graph of a clause represents how chains of substitutions may increase the number of concatenations in the clause.
Consider an edge in the splitting graph from a position $p$ to a position $p'$. 
By definition, there is an intermediate position $p''$ opposite $p$ and carrying the same variable as $p'$.
This means that when splitting decreases the number of concatenations on the side of $p$ by one 
(the label of $p$ may be $y$ referred to in the left split),
the substitution of the label of $p''$ (this would be $x$ in the left split) would cause that the position $p'$ also labeled by $x$ is replaced by the concatenation $x_1\concat x_2$.
Moreover, since the length of the side of $p'$ is now larger, it is possible to perform more splitting steps that follow edges starting at the side of $p'$ and increase numbers of $\concat$ at positions reachable from $p'$ and consequently also further along the path in the splitting graph starting at $(p,p')$. 
Hence the intuitive meaning of the edge is that decreasing the number of $\concat$ at the side of $p$ might increase the number of $\concat$ at the side of $p'$.
Chain-freeness now guarantees that it can happen only finitely many times that decreasing the number of $\concat$ at the side of a position $p$ can later lead to increasing this number through a sequence of splitting steps. 

%\lukas{included from appendix}
%%%%%%%%%%%%%% BEGIN JOURNAL TERMINATION CHAIN-FREE PROOF %%%%%%%%%%%%%%%%%%%%%%%%%%%%%%%%%%%%%%
%\input{sections/termination-proof-lukas}
\section{Proof of Theorem~\ref{theorem:splitting}}
\label{section:termination}

Through the proof, we may occasionally abuse notation and treat clauses $\clause = \constr_1\land\cdots\land\constr_m$ as the sets $\{\constr_1,\ldots,\constr_m\}$.

We first prove a most basic property of splitting, preservation of  chain-freeness.
\begin{lemma} 
\label{lemma:chain-free}
Splitting steps preserve chain-freeness.
\end{lemma}
\begin{proof}
By contradiction. Assume that a clause $\clause'$ is obtained from a chain-free clause $\clause$ by a left split, but it contains a chain. We omit the case of a right split as it is analogous.
Construction of $\clause'$ from $\clause$ might be understood as a sequence of two steps.
The first constructs $\clause[x/x_1\concat x _2]$,
the second replaces the constraint $\relconstrof {x\concat\term} {y\concat\term'}\rel[x/x_1 \concat x_2] $
by the conjunction $\relconstrof {x_1} {y} {R_i}\land \relconstrof {x_2\concat\term} {\term'} {\rel_i'}$ 
(for some $i$, see the definition of the left split).
When reasoning modulo isomorphism of graphs, ignoring the mappings $\var$ and $\edgecons$, then the second step only removes edges from the splitting graph (the positions of the new left conjunct now cannot use the positions of the new right conjunct as the intermediate positions for creating edges and vice versa).
Therefore the chain must have been created during the first step, the substitution, and so it must be present already in $\clause[x/x_1\concat x _2]$. 
Let us define for a position $p = (i,j)$ of $G_{\clause[x/x_1\concat x _2]}$ its \emph{predecessor} as the position $\bar p = (i-k,j)$ of $G_\clause$ where 
$k$ is the number of positions $(i',j),i'\leq i$ with $x_2$. 

We will argue that for each edge $(p_1,p_2)$ of $G_{\clause[x/x_1\concat x _2]}$, $(\bar p_1,\bar p_2)$ is an edge in $G_\clause$, and therefore the presence of the chain in  $G_{\clause[x/x_1\concat x _2]}$ implies a presence of a chain on the predecessor positions in $G_\clause$, which is a contradiction.  
Recall that the presence of the edge $(p_1,p_2)$ in $G_{\clause[x/x_1\concat x _2]}$ is defined as an existence of a position $p_3$ opposite $p_1$ with $\var(p_2) = \var(p_3)$. 
Now, by the definitions of a predecessor and of $G_{\clause[x/x_1\concat x _2]}$, we have that 1) if $p_1$ is opposite $p_3$ in $G_{\clause[x/x_1\concat x _2]}$, then $\bar p_1$ is opposite $\bar p_3$ in $G_{\clause}$; and 2) assuming 
$\var (\bar p_i) = z$ in $G_{\clause[x/x_1\concat x _2]}$, then if $z \neq x_j,j\in\{1,2\}$ then $\var(\bar p_i) = z$ in $G_\clause$; and if if $z=x_j,j\in\{1,2\}$, then $\var(\bar p_i) = x$ in $G_\clause$. 
Hence, if $\var (p_2) = \var (p_3)$ in $G_{\clause[x/x_1\concat x _2]}$, then $\var (\bar p_2) = \var (\bar p_3)$ in $G_{\clause}$.
Therefore $(\bar p_1,\bar p_2)$ is indeed an edge in $G_\clause$.
\end{proof}

Next, we will show that Phase 1 terminates. 
Let $\clause$ be a clause that appears during Phase~1. 
%
%For a root constraint $\rel(t,t')$ of $\reminder$, 
The side of a root constraint of $\reminder$ which consists exclusively of root positions of $G_\reminder$ is called an \emph{outer side},
and a side of a root constraint which is not an outer side (if there is one) is called an \emph{inner side} (note that only a root constraint can have an outer or an inner side and every root constraint has an outer side).
The termination proof will use several auxiliary observations:

\begin{lemma}
\label{lemma:aux1}
For every clause $\clause$ that appears during Phase 1 and for every clause $\clause'$ created from it by splitting, it holds that:
\begin{enumerate}[(1)]
\item \label{lemma:root}
A non-empty chain-free clause has a root constraint.
\item \label{lemma:hroot} 
Every non-empty reminder $\reminder$ created during Phase 1 has a root constraint.
\item \label{lemma:outerside}
A constraint side is an outer side of $\reminder$ if and only if none of its variables has multiple occurrences in $\reminder$.  
\item\label{lemma:alwaysroot}
A successor of a root constraint of $\reminder$ that appears in $\clause'$
is a root constraint of $\reminder'$.
Further, a successor of an outer side in $\reminder$ remains an outer side in $\reminder'$.
\item\label{lemma:outerdown}
A successor of an outer side which is in $\reminder'$ is at most as long as the outer side.
\item\label{lemma:innerorouter}
Either (a) the successor in $\reminder'$ of the outer side of the split constraint is strictly shorter that the original outer side, or (b) the successor of its opposite side is strictly shorter and, at the same time, the successors of all the other constraint in $\reminder$ are the same as the originals.   
%
%and it if it is in the split constraint, then either its successor is shorter or the  is strictly shorter if its constraint is the split constraint. 
\end{enumerate}
\end{lemma}

\begin{proof}
\begin{itemize}
\item[(\ref{lemma:root})]
By contradiction. Assume that there is no root constraint in the chain-free clause. 
That means that in the splitting graph, each side of a constraint has an edge  leading from it. 
Define high-level edges between sides of constraints such that there is an edge from $s$ to $s'$ if and only if there is an edge in the splitting graph from a position inside $s$ to a position inside $s'$.
By our initial assumption, every constraint side has an incoming high-level edge, which implies that there is a cycle $(s_0,s_1),(s_1,s_2),\ldots,(s_{n-1},s_n),s_n = s_0$ constructed from the high-level edges. 
We will show that this cycle induces a chain in the splitting graph. 
By the definition of the high-level edges and of the splitting graph, 
the existence of each high-level edge $(s_i,s_{i+1})$, $0\leq i < n$, implies that there are three positions $p_i,p''_i,p'_i$ such that $e_i = (p_i,p_i')$ is an edge of the splitting graph, $p_i$ is a position of $s_i$, $p_i'$ is a position of $s_{i+1}$, and $p''_i$ is opposing $p_i$ and has the same label as $p_i'$. 
Because $p_{(i-1) \modulo ({n+1})}'$
is also a position of $s_i$, and hence opposing $p_i''$, 
the splitting graph has also the edge 
$e_i' = (p_{(i-1) \modulo ({n+1})}',p_i')$, 
end the edges $e_1',\ldots,e_n'$ are a chain in the splitting graph.
This contradicts the assumption that the splitting graph is chain-free.
\item[(\ref{lemma:hroot})]
As a subset of a chain-free clause $\clause$, $\reminder$ must also be chain-free, and hence, by (\ref{lemma:root}), it has a root constraint. 
\item[(\ref{lemma:outerside})]
By the definitions of an outer side and a splitting graph. A position has an incoming edge if and only if its variable occurs multiple times in $G_{\reminder}$. 
\item[(\ref{lemma:alwaysroot})]
A position of an outer side might be the split position, it may be opposite the split position, or neither.  
By definition of a splitting step and because of (\ref{lemma:outerside}),  
in all the three cases, the variable $z$ on the split position, as well as each of its two variants $z_1$ and $z_2$ that might have been created by the split, may appear only once in the successors of the constraints of $\reminder$ (The case of $z$ on the split position creates two constraint. One concatenation-free with the outer side $z_1$, the other one with $z$ replaced by $z_2$. The other two cases do not substitute $z$.) 
Since $\reminder'$ is a subset of successors of $\reminder$, the $z$ appears at most once there as well.
\item[(\ref{lemma:outerdown})]
Due to (\ref{lemma:outerside}), 
the outer side does not change when another constraint is split, hence its successor has the same length.   
If the constraint of the outer side was the split constraint, then there are two possibilities: 

a by) The outer side contains the split position, hence, one successor is obtained by replacing the left-most variable $z$ by $z_2$ and it is of the same length, while the other successor is $z_1$. The successor $z_1$ a concatenation-free constraint, it is also a root constraint by (\ref{lemma:alwaysroot}), and hence it is removed from $\reminder'$ by definition of Phase 1. 

b) The split position is opposite the outer side, in which case the successor of the outer side is obtained by removing the left most variable of the original, and hence it is shorter 
(and is removed from $\reminder'$ in case it is concatenation-free since it is still a root constraint by (\ref{lemma:alwaysroot})). 
%, or it is the left-most variable but also concatenation-free and thus not present in $\reminder'$ by definition of splitting step. 

\item[(\ref{lemma:innerorouter})]
%Continuing the argument from (\ref{lemma:outerdown}):
The successor of the outer side of the split constraint cannot be longer by (\ref{lemma:outerdown}). 
If (a) the split position is on the side opposite the outer side, then the successor of the outer side is shorter, by the definition split and because of by  (\ref{lemma:outerside}), the variable on the split position does not occur in the outer side. 
If (b) the outer side contains the split position, then its successor is of the same length by the definition split and by (\ref{lemma:outerside}). 
(\ref{lemma:outerside}) is also the reason for that no other constraint is influenced by the substitution.  
\end{itemize}%
\end{proof}

With Lemma~\ref{lemma:aux1} at hand, we are ready to show termination of Phase~1.

\begin{lemma}
Phase~1 terminates.
\end{lemma}

%\begin{lemma}
%\label{lemma:aux1}
%For every clause $\clause$ that appears during Phase 1 and for every clause $\clause'$ created from it by splitting, it holds that:
%\begin{enumerate}[(1)]
%\item \label{lemma:root}
%A non-empty chain-free clause has a root constraint.
%\item \label{lemma:hroot} 
%Every non-empty reminder created duting Phase 1 has a root constraint.
%\item \label{lemma:outerside}
%A constraint side is an outer side of $\reminder$ if and only if each its variable has only single occurence in $\reminder$.  
%\item\label{lemma:alwaysroot}
%A successor of a root constraint of $\reminder$ that appears in $\clause'$
%is a root constraint of $\reminder'$.
%Further, a successor of an outer side in $\reminder$ remains and outer side in $\reminder'$.
%\item\label{lemma:outerdown}
%A successor of an outer side which is in $\reminder'$ is not longer that the outer side.
%\item\label{lemma:innerorouter}
%Either (a) the successor of the outer side of the split constraint in $\reminder'$ is strictly shorter, or (2) the successor of its opposite side is strictly shorter and, at the same time, all the other constraint in $\reminder$ have unchanged successors.  
%\end{enumerate}
%\end{lemma}

\begin{proof}
Consider a sequence of clauses $\clause_0,\clause_1,\ldots$  
such that $\clause_0$ is an initial clause and $\clause_i$ is created from $\clause_0$ in Phase~1 by a splitting step. To show that the sequence terminates,
we annotate each step $j$ in the sequence with a measure $m(j) = (l,o,i)$ where 
\begin{itemize}
\item
$l$ equals the number of the original constraints minus the number of times a successor of a non-root constraint becomes a root constraint,
\item
$o$ equals the sum of the lengths of the outer sides of the roots the constraints in $\reminder$, and
\item
$i$ equals the sum of the lengths of the inner sides of the roots the constraints in $\reminder$.
\end{itemize}
We define $(l,o,i)<(l',o',i')$ to hold if and only if
$$
(l < l') \lor (l = l' \land o < o') \lor (l = l' \land o = o' \land i < i')\ .
$$
%Notice that $0\leq l \leq n_0$, where $n_0$ is the number of the original constraints in $\clause$, is an invariant of Phase 1 
%(i.e., at most as many times a root constraint is created as there are constraints in the original clause). 
Notice that 1) a splitting step cannot increase the number of constraints in the reminder graph because every constraint has at most one successor in it 
(one of the two constraints of the disjunct of the split is concatenation-free and hence not in the reminder graph),
2) the successor of a root constraint either remains a root constraint or is removed from the reminder graph (Lemma~\ref{lemma:aux1} (\ref{lemma:alwaysroot})). %introducing one new root constraint means decreasing the number of constraint in the reminder graph, and
Therefore, for all clauses created in Phase~1, the first coordinate of $m(j)$ cannot be smaller than 0. The second and the third coordinate is a natural number by definition. $m(j)$ is hence bounded from below by $(0,0,0)$.  
We can therefore show the termination of Phase~1 by showing that the measure is strictly decreasing with each splitting step, that is, for a clause $\clause_j$ created from $\clause_{j-1}$ by a splitting step, $m(j) < m({j-1})$.

%If a constraint from the reminder graph became a root constraint, 
%then $l > l'$ and we are done.
%The case $l < l'$ can never happen by definition of $l$ and since the number of constraints in the reminder graph is never increasing.
%%
%If $l = l'$ and some outer side got shorter, then we have $o > o'$ and we are done. 
%Since $l = l'$, the set of roots is the same as before.  Hence it is not possible that $o < o'$ because the variables in outer sides have only a single occurrence in the reminder graph (because outer sides consist of positions that are leaves of the reminder graph), and no constraint side can get longer by splitting at a variable with a single occurrence. %substituting such variable by splitting of constraint in the reminder graph (be definition of split).

Let $m(j) = (l,o,i)$ and $m_{j+1} = (l',o',i')$.
If a non-root constraint from the reminder graph became a root constraint, 
then $l > l'$ and we are done (by Lemma~\ref{lemma:aux1} (\ref{lemma:alwaysroot}), a successor of a root constraints in $\reminder_j$ is a root constraint, and the number of constraints does not increase during Phase 1).
The case $l < l'$ cannot happen by definition of $l$ and since the number of constraints in the reminder graph does not increase.
If $l = l'$, then every root constraint in $\reminder_j$ has a successor in $\reminder_{j-1}$. 
By Lemma~\ref{lemma:aux1} (\ref{lemma:outerdown}), an outer sides of the successors are not longer then the originals, hence $o\geq o'$. If $o>o'$, then we are done.
If $o=o'$, then by Lemma~\ref{lemma:aux1} (\ref{lemma:innerorouter}), some inner side of a root constraint got shorter and the other constraints did not change, hence $i>i'$, and so, indeed, $m(j)>m(j+1)$.

Every sequence of splitting thus terminates. 
It remains to account for that each splitting step produces not one but multiple clauses. Phase 1 therefore generates trees of clauses, every generated tree rooted by a clause of the original DNF, and children of each clause generated by a splitting step.  
By the reasoning above, every tree branch is finite.
To show that the entire trees are finite, it remains to argue that no node has infinitely many children (the finiteness of the tree then follows by the K\"onig's lemma).  
This is quite obvious from the definition of a split.  
Every splitting steps creates at most as many clauses as there are disjuncts in the left and the right split.
The number of disjuncts in a split of a transducer constraint is bound by the number of states of the transducer. The number of disjuncts created from an equation is just two.
%, and the resulting disjuncts feature a transducer with the same number of states. An equality produces a single dusjunct for each split. 
%The number of the disjuncts is therefore bounded by twice the number of states of the largest transducer in the original formula, or it is equal to two if there are no transducer constraints.  
%
%
%of the root contraint which was split shorter outer side, and the other outer sides stay the same since their variables cannot occur anywhere else in the reminder graph, by the definition of an outer side. 
%Hence we have $o > o'$. 
%
%If $l = l'$ and $o = o'$, then the set of leaves is the same, but this means that $i$ must have decreased by Lemma~\ref{lemma:aux1}, hence we have that $i > i'$. This concludes the proof of termination of Phase~1.
\end{proof}

Next, we will show that Phase~2 also terminates.
Our argument is based on that clauses returned by Phase~1 have certain special properties. 
We note that these properties imply that the clauses fall into the straight line fragment of \cite{BL16}.
However, we do not use the algorithm from \cite{BL16}. 
The splitting in \cite{BL16} must be done in certain order. 
Our algorithm is more liberal, allowing any order of splitting in Phase 2.
We only use an \emph{existence} of certain ordering of the clauses output from Phase 1 to prove termination Phase 2. 

%Our algorithm allows to apply splitting in any order, which is much more liberal than in \cite{BL16}.

Let $\clauseone$ be a clause at the input of Phase 2, equal to the last clause $\clause_n$ of a sequence of clauses 
$\clause_0,\ldots,\clause_n$ generated in Phase 1 where $\clause_0$ is in the DNF on the input of Phase 1 and $\clause_i$ is created by splitting of $\clause_{i-1}$. 
Let $\drop_i,1\leq i \leq n$ be the set of concatenation-free successors of the root constraint in $\reminder_{i-1}$, those that 
were by definition of Phase 1 not included in $\reminder_i$.
%were split by the $i$th splitting step, and were, being concatenation-free, dropped from $\reminder_{i}$ 
%($\drop_i$ contains one or two elements, by the definition of split). 
%
%
We may now define, for each $i:1\leq i \leq n$,
$\clauseone(i)$ as the set of the constraints of $\clauseone$ that are descendants of constraints in $\drop_i$. 
%
%Further, let the side of a constraint in $\clause(i)$ that is  a successor of the outer side of the predecessor constraint in $\drop_i$ be called outer side.
%
Further, let us call all constraints in $\clauseone(i)$ that all successors of constraints that were outer sides during Phase 1 still outer sides (even though the original notion of outer side was relevant only in the context of a reminder graph). 
%the side of a constraint in $\clause(i)$ that is a successor of a side of a constraint that was once an outer constraint during Phase 1 be called outer side.

To prove termination of Phase~2, we first prove several properties of the clauses output from Phase 1. They are stated in \cref{lemma:init}. Later, in \cref{lemma:inv}, we will generalise them to an invariant of Phase 2. 
\begin{lemma}[Initial conditions]
\label{lemma:init}
\begin{enumerate}[(1)]
\item\label{lemma:cover}
$\clauseone = \bigcup_{0\leq i \leq n}\clauseone(i)$
\item\label{lemma:singleton}
The outer side of every constraint of $\clauseone$ is a single variable. 
\item\label{lemma:notupwards}
An outer side variable of a constraint in $\clauseone(i)$ (a) cannot appear in $\clauseone(j)$ for $j>i$ 
and (b) cannot appear twice in $\clauseone(i)$.
\item\label{lemma:distinct}
Outer sides in $\clauseone$ are pairwise different.
%Each outer side variable of $\clauseone$ has only single occurrence in $\clauseone$.
\end{enumerate}
\end{lemma}

\begin{proof}
\begin{itemize}
\item[(\ref{lemma:cover})]
That $\clauseone \supseteq \bigcup_{0\leq i \leq n}\clauseone(i)$ is trivial since every $\clauseone(i)$ is a subset of $\clauseone$ by definition. 
To show the other inclusion, 
recall that every constraint $\constr$ of $\clauseone$ is reached from the original clause $\clause_0$ as the end of a sequence of successors generated in Phase 1. The sequence ends in a constraint that does not belong to any reminder seen along generating the sequence in Phase 1 since $\reminder_n$ is empty. Hence, for some $i:1\leq i \leq n$, the $i$th successor in the sequence is the first one which does not appear in the reminder $\reminder_i$. It is therefore in $\drop_i$ and hence $\constr$ is in $\clauseone(i)$ by definition of $\clauseone(i)$.
%
%Phase 1 ends when the reminder graph is empty.  end by the reminder graph being empty and constraints that a
%Phase 1 starts with the reminder graph containing all constraints of the original formula.
%All constraints of $\clause$ are successors of the constraints of the original formula (they are created by splitting from the original formula).
%When a successor of a constraint of the original formula is removed from reminder graphs during Phase 1, it is added to some $\drop(i)$ and its successor will appear in $\clause(i)$. 
%Phase 1 ends with the reminder graph empty, so every constraint in $\clause$ has a predecessor that was removed from reminder graphs and hence it belongs to some $\clause(i)$.
\item[(\ref{lemma:singleton})]
When a constraint is added to $\drop_i$, it is concatenation-free, so its outer side is a singleton.
Since it is an outer side, its variable $z$ does not appear in the reminder $\reminder_i$.
It cannot hence be affected by the remaining part of Phase 1, 
because the reminder of Phase 1 is only concerned with variables in the reminder $\reminder_i$.  
%The outer side thus stays $z$ until the end of Phase 1.
\item[(\ref{lemma:notupwards})]
(a) As we argued above, after addition to $\drop_i$, variables of outer sides of constraints do not take part in the remaining steps of Phase 1. 
They also cannot appear in any $\reminder_j$ with $j>i$ since each reminder only gets a subset of the  successors of the constraints in the previous reminder.
Hence the variables of the outer sides in $\drop_i$ cannot appear in $\drop_j$ with $j>i$.  

(b) That the outer side variable from $\clauseone(i)$, say $z$, cannot appear twice in $\clauseone(i)$ follows from that the constraints in $\drop_i$ are root contraints (by the definition of Phase 1 and \cref{lemma:aux1}.(\ref{lemma:alwaysroot})), and from that by \cref{lemma:aux1}.(\ref{lemma:outerside}), $z$ does not occur multiple times in the reminder $\reminder_i$ (substitution in the later steps of Phase 1 then cannot create multiple occurences of a single variable from the single occurence of  $z$).

%successors of a single root constraint $\constr$\lukas{fix!} and the outer sides in $\clause_(i)$ are descendants of an outer side of $\constr$. 
%Outer sides of $\constr$ cannot have multiple occurrences of variables by \cref{lemma:outerside}  of \cref{lemma:aux1}. 
%Their descendants are created by splitting and that cannot generate multiple occurrences of a variable from a single occurrence of a variable.

%At that point, the variable $z$ at the outer side does not appear in the reminder graph, since it is an outer side of the leaf. It will never propagated to other constraints during Phase 1 because further splitting takes place only inside reminder graphs.  
%Constraints in $\clause(j),j> i$ are successors of constraints that are after the $i$th split still in the reminder graph. Hence these predecessors of the of the constraints in $\clause(j)$ cannot contain $z$, and hence the constraints themselves cannot contain $z$ too (since $z$ does not appear in any further splitting in Phase 1). 
%\item[(\ref{lemma:distinct})]
This follows directly from the previous points (\ref{lemma:cover}), (\ref{lemma:singleton}), and (\ref{lemma:notupwards}).
\end{itemize}
\end{proof}

Further, let $\clauseone = \clauseone_0,\clauseone_1,\ldots$ be a sequence of clauses created each from the previous by a splitting step during Phase 2.
We will translate the partition of constraints in $\clause$ into $\bigcup_{0\leq i \leq n}\clauseone(i)$ into Phase 2 by defining
$\clauseone_0(i) \defeq \clause(i) $ and 
defining $\clauseone_{j+1}(i)$ as the set of the successors of the constraints in $\clauseone_{j}(i)$, for all $0\leq i \leq n$.
This allows us to generalise Lemma~\ref{lemma:init} as an invariant of Phase 2: 

\begin{lemma}[Invariant] For all $j\geq 0$ it holds that
\label{lemma:inv}
\begin{enumerate}
\item\label{lemma:cover2}
%$\{\clauseone_j(i)\mid 0\leq i \leq n\}$  covers the constraints of $\clauseone_j$.
$\clauseone_j = \bigcup_{0\leq i \leq m}\clauseone_j(i)$
\item\label{lemma:singleton2}
The outer side of every constraint of $\clauseone_j$ is a singleton.
\item\label{lemma:notupwards2}
An outer side (a single variable by the previous point) of a constraint in $\clauseone_j(i)$ cannot appear in a constraint in $\clauseone_j(k)$ for $k>i$ 
and cannot appear twice in $\clauseone(i)$.
\item\label{lemma:distinct2}
Outer side variables of $\clause_j$ are pairwise different.
%Each outer side variable of $\clauseone_j$ has only single occurrence in $\clauseone_j$.
\end{enumerate}
\end{lemma}
\begin{proof}
By induction to $j$. All the four points hold initially for $j = 0$ due to Lemma~\ref{lemma:init}. 
Assume now that they hold for some $j$. We will show that they hold after the splitting step for $j+1$. 
\begin{itemize}
\item[\ref{lemma:cover2}.]
%Because 1. holds for $\clauseone_j$ and because for each $i$ s.t. $0\leq i < n$, $\clauseone_{j+1}(i)$ is by definition the set of successors of $\clauseone_j(i)$ and the entire $\clauseone_{j+1}$ is the set of successors of $\clauseone_j$.
Because 1. holds for $\clauseone_j$ and because for each $i$ s.t. $0\leq i < n$, $\clauseone_{j+1}(i)$ is by definition the set of successors of $\clauseone_j(i)$ and the entire $\clauseone_{j+1}$ is the set of successors of $\clauseone_j$.
\item[\ref{lemma:singleton2}.]
The condition is preserved through splitting since when splitting a constraint $\constr$ with the outer side being a singleton $z$, the variable $z$ must be the one being substituted by the concatenation of a two fresh variables $z_1,z_2$, by the definition of split. But since point \ref{lemma:distinct2} was true for $j$, this substitution could not affect any other outer side in $\clauseone_j$. 
The splitting on $\varphi$ produces two constraints, one with the outer side $z_1$ and the other with the outer side $z_2$ (both singletons).  
\item[\ref{lemma:notupwards2}.]
Let $\constr \in \clauseone_j(i)$ have an outer side $z$. By (\ref{lemma:singleton2}) being true for $j$ (ind. hypothesis) and by the definition of split, $\constr$ is split into two constraints with the outer sides $z_1$ and $z_2$ that belong to $\clauseone_{j+1}(i)$. At the same time, no other outer side of is influenced by the split due to that (\ref{lemma:distinct2}) holds for $j$ (ind. hypothesis). Therefore, since  (\ref{lemma:notupwards2}) was true for $j$ (ind. hypothesis), it continues to be true for $j+1$.
%
%There are three possibilities according to what kind of split happens. 1)  It replaces $\constr$ with two constraints with outer sides $z_1$ and $z_2$,  both in $ $  only replace a constraint with an outer side in $\clauseone_{j-1}(i)$ can be only replaced by two constraints with fresh variables on the outer sides in $\clauseone_{j}(i)$, point \ref{lemma:notupwards2} holds.
\item[\ref{lemma:distinct2}.]
Directly implied by (\ref{lemma:cover2}), (\ref{lemma:singleton2}), and (\ref{lemma:notupwards2}).
%\item[\ref{lemma:distinct2}.]
%The only way a splitting step can create new positions is by substitution. By the induction hypothesis, (\ref{lemma:distinct2}) holds for $j$. Substitution cannot generate more occurrences of a variable by substitution from a single occurrence. 
\end{itemize}
\end{proof}

With the lemma above at hand, we can finish the proof.
\begin{lemma}
Phase 2 terminates.
\end{lemma}

\begin{proof}
We will use the decreasing well founded measure defined as follows.
For $j\geq 0$,  $m(\clauseone_j)$ is the vector $(\iota_0,\ldots,\iota_n)$ where $\iota_k$ is the sum of the lengths of the inner sides of constraints in $\clauseone_j(k)$ that are not concatenation-free.  
The ordering $<$ on vectors is the usual lexicographic ordering (the $n$th position is the most significant).
Obviously, $m(\clauseone_j)\geq 0^n$ for all $j$.
We will argue that $m(\clauseone_j)>m(\clauseone_{j+1})$ for all $j\geq 0$:

Assume that the splitting step has split a constraint in $\clauseone_j(k)$.
By the invariant expressed by \cref{lemma:inv}, 
the substitution in the split can affect (and increase their length) only of inner sides of constraints at the level strictly higher than $j$. 
The sum of the lenghts of the inner sides that are not concatenation-free at the levels smaller than $j$ hence does not change. 
At the level $j$, by the definition of split, a constraint with the inner side of the length $\ell$ is replaced by two constraints with the inner sides of the length $1$ and $\ell-1$. The former one is concatenation-free.
The sum of the lenghts of the inner sides that are not concatenation-free at the level $j$ therefore decreases at least by $1$.
This means that $m(\clauseone_j)>m(\clauseone_{j+1})$.
\end{proof}

Let us summarise the proof.
Since Phase 1 and Phase 2 terminate, the algorithm terminates. Since it only uses splitting to modify the formula and since by definition, splitting produces an equisatisfiable formula, the result is equisatisfiable.
Since splitting preserves chain-freeness, the resulting formula is chain-free. 
Since the only termination condition of Phase 2 is concatenation-freeness and Phase 2 terminates, the resulting formula is concatenation-free. This concludes the proof of Theorem~\ref{theorem:splitting}. 
%%%%%%%%%%%%%% END JOURNAL TERMINATION CHAIN-FREE PROOF %%%%%%%%%%%%%%%%%%%%%%%%%%%%%%%%%%%%%%

\section{Satisfiability of Chain and Concatenation-Free Formula}
\label{sec:concat-free}

In this section, we explain an algorithm that decides satisfiability of a chain-free and concatenation-free formula.
The algorithm is essentially a combination of two standard techniques.
First,  concatenation and chain-free conjunction over relational constraints 
is a formula in the "acyclic fragment" of \cite{BFL13} that can be turned into a single equivalent transducer constraint (an approach used also in e.g. \cite{sloth}).
Second, consistency of the resulting transducer with the input length constraints may be checked via computation of the Parikh image of the transducer.

We will now describe the two steps in a more detail.
For simplicity, we will assume only transducer and length constraints. This is without loss of generality because the other types of constraints can be encoded to transducers.

\paragraph{\bf Transducer constraints.}
A conjunction of transducer constraints may be decided 
through computing an equisatisfiable  multi-tape transducer constraint and checking emptiness of its language. 
The transducer constraint is computed by synchronizing pairs of constraints in the conjunction.
That is, 
synchronization of two transducer constraints 
$\transducer_1(x_1,\ldots,x_n)$ and $\transducer_2(y_1,\ldots,y_m)$ is possible if they share at most one variable
(essentially the standard automata product construction where the two transducers synchronise on the common variable).
The result of their synchronization is then a constraint 
$\transducer_1 \wedge_{(i,j)} \transducer_2(x_1,\ldots,x_n,y_1\ldots y_{j-1},y_{j+1},\ldots,y_m)$ 
where $y_j$ is the common variable equal to $x_i$ for some $1\leq i \leq n$
or a constraint $\transducer_1 \wedge \transducer_2(x_1,\ldots,x_n,y_1,\ldots, y_m)$ if there is no common variable. The $\transducer_1 \wedge \transducer_2$ is a loose version of $\wedge_{(i,j)}$ that does not synchronise the two transition relations  
(see e.g. \cite{sloth,BL16} for details on implementation of a similar construction).
Since the original constraint is chain and concatenation-free, two constraints may share at most one variable.
This property stays an invariant under synchronization steps, and so they may be preformed in any order until only single constraint remains. 
Termination of this procedure is immediate because every step decreases the number of constraints.

\medskip

\noindent
{\bf Length constraints.}
\newcommand{\symbvect}{\alpha}
\newcommand{\symbi}{a_i}
\newcommand{\parikhf}{\formula_{\mathrm{Parikh}}}
\newcommand{\alphaf}{\Phi}
\newcommand{\vectvars}{\mathbb{P}}
\newcommand{\pairvars}{\mathbb{A}}
A formula of the form $\formula_r\land\formula_l$ where $\formula_r$ is a conjunction of relational constraints and $\formula_l$ is a conjunction of length constraints
may be decided through replacing $\formula_r$ by an existential Presburger formula $\formula_r'$ over length constraints that captures the length constraints implied by $\formula_r$. 
That is, an assignment $\nu:\{|x|\mid x\in\varset\}\rightarrow \nat$ is a solution of $\formula_r'$
if and only if there is a solution $\eta$ of $\formula_r$ such that $|\eta(x)| = \nu(|x|)$ for all $x\in \varset$.
The conjunction $\formula_r'\land\formula_l$ is then an existential Presburger formula equisatisfiable to the original conjunction, solvable by an of-the-shelf SMT solver.

Construction of $\formula_r'$ is based on computation of the Parikh image of the synchronised constraint $\transducer(x_1,\ldots,x_n)$ equivalent to $\formula_r$. 
Since $\transducer$ is a standard finite automaton over the alphabet of $n$-tuples $\alphabet_\emptystr^n$, its Parikh image can be computed in the form of a semi-linear set represented as an existential Presburger formula $\parikhf$ by a standard automata construction (see e.g. \cite{parikh}). 
The formula captures the relationship between the numbers of occurrences of letters of $\alphabete^n$ in words of $\langof\transducer$. 
Particularly, the numbers of letter occurrences are represented by the \emph{Parikh variables} $\vectvars = \{\#\symbvect\mid \symbvect \in \alphabete^n\}$ and it holds that $\nu\models\parikhf$ iff there is a word $w\in \langof\transducer$ such that for all $\symbvect\in\alphabete^n$, $\symbvect$ appears $\nu(\#\symbvect)$ \mbox{times in $w$.}

The formula $\formula_r'$ is then extracted from $\parikhf$ as follows.
Let $\pairvars = \{\#\symbi \mid a\in\alphabet,1\leq i \leq n\}$ be a set of \emph{auxiliary variables} 
expressing how many times the letter $a\in \Sigma$ appears on the $i$th position of a symbol from $\alphabete^n$ in a word from   $\langof\transducer$. 
Let $\symbvect[i]$ denotes the $i$th component of the tuple $\symbvect\in\alphabete^n$.
We construct the formula $\alphaf$ that uses variables $\pairvars$ to describe the relation between values of 
$|x_1|,\ldots,|x_n|$ and variables of $\vectvars$:
$$
\alphaf :=
\bigwedge\nolimits_{i=1}^n 
\left( 
	|x_i| = \sum\nolimits_{a\in\alphabet} \#\symbi  
	\land 
	\bigwedge\nolimits_{a\in\alphabet} 
	\Bigl( 
		\#\symbi = 
		\sum\nolimits_{\symbvect \in \alphabete^n \text{ s.t. } \symbvect[i] = a}
		\#\symbvect
	\Bigr)
\right) 
$$
%
%We the obtain $\formula_r'$ by eliminating the quantifiers from 
%$\exists \vectvars \exists \pairvars :\alphaf\land\parikhf$.
%
We then obtain $\formula_r'$ by in the form 
$\exists \vectvars \exists \pairvars :\alphaf\land\parikhf$.

\section{Experimental Results}
\label{experiments:section}

\newcommand{\maintable}{
\newcolumntype{g}{>{\columncolor{white!20}}l}
\renewcommand{\arraystretch}{0.8}
\begin{table*}[htbp]
\centering
\normalsize{
\begin{tabular}{|g|l|>{\centering\arraybackslash}p{1.3cm}|>{\centering\arraybackslash}p{1.3cm}|>{\centering\arraybackslash}p{1.3cm}|>{\centering\arraybackslash}p{1.3cm}|>{\centering\arraybackslash}p{1.3cm}|>{\centering\arraybackslash}p{1.3cm}|}
\hline
\rowcolor{white}						&				& Ostrich			& Z3-str3	& CVC4		& \Trau 	& \Trauplus \\ \hline	
\rowcolor{gray!20}\cellcolor{white}		&	sat        	& 0             		& -		& -				& - 	 	& {\bf 5}\\ 
\rowcolor{gray!20}\cellcolor{white}		&	unsat		& 0             		& -		& -     		& - 	 	& {\bf 14}\\ 
										&	timeout     & 6              	& -		& -        		& -    		& 7 \\  
\multirow{-4}{*}{\parbox{21mm}{{\bf Chain-Free} \\[-0mm] (26)}}	& error/unkn. & 20              	& -		& -          	& -     	& 0 \\ \hline
\rowcolor{gray!20}\cellcolor{white}		&	sat        	& {\bf 106}			& -		& -				& 26     	& 105 \\ 
\rowcolor{gray!20}\cellcolor{white}		&	unsat		& {\bf 14}	   		& -		& -				& 4 	    & {\bf 14} \\ 
										&	timeout     & 0		            & -		& - 			& 0	     	& 1 \\  
\multirow{-4}{*}{\parbox{15mm}{{\bf ReplaceAll}\\[-0mm] (120)}}	& error/unkn. & 0		            & -		& - 			& 90   		& 0 \\ \hline  
\rowcolor{gray!20}\cellcolor{white}		&	sat        	& 1250             	& 298	& 1278			& 1174 	 	& {\bf 1287}\\ 
\rowcolor{gray!20}\cellcolor{white}		&	unsat		& 2022             	& 2075	& 2079 	 		& 2080 	 	& {\bf 2081}\\ 
										&	timeout     & 3              	& 903	& 9        		& 24    	& 23 \\  
\multirow{-4}{*}{\parbox{15mm}{{\bf Replace} \\[-0mm] (3392)}}	& error/unkn. & 117              	& 116	& 26          	& 114     	& 1 \\ \hline  
\rowcolor{gray!20}\cellcolor{white}		&	sat        	& 36					& 839	& 4178			& 4244     	& {\bf 4245} \\ 
\rowcolor{gray!20}\cellcolor{white}		&	unsat		& 299       			& 1477		& 1281 			& 1287      	& {\bf 1287} \\ 
										&	timeout     & 0		            & 3027		& 105  			& 35	     	& 35 \\  
\multirow{-4}{*}{\parbox{15mm}{{\bf PyEx-td} \\[-0mm] (5569)}}	& error/unkn. & 5234              & 226		& 5  			& 3	   			& 2 \\ \hline  
\rowcolor{gray!20}\cellcolor{white}		&	sat        	& 35             	& 1211		& 5617			& 6680     	 	& {\bf 6681}\\ 
\rowcolor{gray!20}\cellcolor{white}		&	unsat		& 466             	& {\bf 1870}& 1346     		& 1357     	 	& 1357\\ 
										&	timeout     & 0              	& 4760		& 1449         	& 374	    	& 374\\  
\multirow{-4}{*}{\parbox{15mm}{{\bf PyEx-z3} \\[-0mm] (8414)}}	& error/unkn. & 7913              & 573		& 2           	& 3	     	 	& 2\\ \hline  
\rowcolor{gray!20}\cellcolor{white}		&	sat        	& 38		        	& 2840		& {\bf 9817}	& 8966      	& 8967\\ 
\rowcolor{gray!20}\cellcolor{white}		&	unsat		& 141           	  	& 1974		& {\bf 1202}	& 1192     	 	& 1193\\ 
										&	timeout     & 0              	& 5988			& 416          	& 1277	    	& 1276\\  
\multirow{-4}{*}{\parbox{15mm}{{\bf PyEx-zz} \\[-0mm] (11438)}}	& error/unkn. & 11259            	& 636			& 3           	& 3	     		& 2\\ \hline \hline   
\rowcolor{gray!20}\cellcolor{white}		&	solved      	& 4407             	& 12730			& 26798			& 27010    	& {\bf 27236}\\ 
\multirow{-2}{*}{\parbox{15mm}{{\bf Total} \\[-0mm] (28959)}}	&	unsolved		& 24552            	& 16229			& 2161     		& 1949     	& {\bf 1723}\\ \hline  
\end{tabular}
}

\caption{
\normalsize{Results of running solvers over Chain-Free, two sets of the SLOG, and four sets of PyEx suite.}
} 
%\vspace*{-8mm}
\label{fig:results-all} 
\end{table*}
}

We have implemented our decision procedure in {\sc Sloth} \cite{sloth} and then used it in the over-approximation module of the string solver \Trauplus.
 \Trauplusspace  is as an open source string solver, it is an extension of \Traunospace~\cite{trau18}, and uses Z3 \cite{DeMoura08Z3} as the SMT solver to handle generated arithmetic constraints.
\Trauplusspace is based on  a Counter-Example Guided Abstraction Refinement (CEGAR) framework which contains both an under- and an over-approximation module. These two modules interact together in order to automatically  make these approximations more precise. The extension of  {\sc Sloth} in the over-approximation module of \Trauplusspace takes as an input a constraint and checks if it belongs to the weakly-chaining fragment. If it is the case, then we use our decision procedure outlined above. Otherwise, we start  by  choosing a minimal set of  occurrences of  variables $\xvar$ that need to be  replaced by  fresh ones so that the resulting constraint falls in our decidable fragment. 

We compare \Trauplusspace performance  against the performance of  four other state-of-the-art string solvers, namely Ostrich \cite{Chen:2019}, Z3-str3 \cite{Berzish2017Z3str3AS}, CVC4 1.6 \cite{LiaEtAl-CAV-14,CVC4}, and  {\sc Trau}  \cite{Trau}. For our comparison with Z3-str3, we use the version that is a part of Z3 4.8.4. The goal of our experiments is twofold:

\begin{itemize}
	\item \Trauplusspace handles transducer constraints in an efficient manner. It can handle more cases than 
%\Traunospace since  the new over-approximation of \Trauplusspace supports more and new  transducer constraints  that the one of 
\Traunospace.

	\item \Trauplusspace performs either better or as well as existing tools on transducer-less benchmarks.
\end{itemize}

We carry experiments on benchmark suites that draw from the real world applications with diverse characteristics. The first suite is our new suite Chain-Free. Chain-Free is obtained from variations of various PHP codes, including the introductory example. The second suite is SLOG \cite{DBLP:conf/cav/WangTLYJ16} that is derived from security analysis of real web applications. The suite was generated by the authors of Ostrich. The last suite is PyEx \cite{CVC417}, derived from PyEx - a symbolic executor designed to assist Python developers to achieve high coverage testing. The suite was generated by CVC4 group on 4 popular Python packages {\tt httplib2}, {\tt pip}, {\tt pymongo}, and {\tt requests}. The summary of the main experimenting with Chain-Free, SLOG, and PyEx is given in Table \ref{fig:results-all}. All experiments were performed on an Intel Core i7 2.7Ghz with 16 GB of RAM.
The time limit is 30s for each test which is widely used in the evaluation of other string solvers. Additionally, we use 2700s for Chain-Free suite, much larger than usual as its constraints are difficult. Rows with the heading ``sat''/``unsat'' indicate the number of times the solver returned satisfiable/unsatisfiable. Rows with the heading ``timeout'' indicate the number of times the solver  exceeded the time limit. Rows with the heading ``error/unknown'' indicate the number of times the solver either crashed or returned unknown. 

\maintable

Chain-Free suite consists of 26 challenging chain-free tests, 6 of them being also straight-line. The tests contain Concatenation, ReplaceAll, and general transducers constraints encoding various JavaScript and PHP functions such as htmlescape, escapeString. Since Z3-str3, CVC4, and \Trau do not support the language of general transducers, we skip performing experiments on those tools in the suite. Ostrich returns 6 times ``timeout'' for straight-line tests and times 20 ``unknown'' for the rest. \Trauplusspace handles well most cases, and gets ``timeout'' for only 7 tests.

SLOG suite consists of 3512 tests which contain transducer constraints such as Replace and ReplaceAll. Since Z3-str3 and CVC4 do not support the ReplaceAll function, we skip doing experiments on those tools in the ReplaceAll set. In both sets, the result shows that \Trauplusspace clearly improved \Traunospace. In particular, \Trauplusspace can handle most cases where \Trau returns either ``unknown'' and ``timeout''. \Trauplusspace has also better performance than other solvers.

PyEx suite consists of  25421 tests which contain diverse string constraints such as IndexOf, CharAt, SubString, Concatenation. \Trau and CVC4 have similar performance on the suite. While 
 \Trau is better on PyEx-dt and PyEx-z3 sets (3 less error/unknown results, roughly 1000 less timeouts),
CVC4 is better on PyEx-zz set (about 800 less timeouts). CVC4 and \Trau clearly have an edge over Z3-str3 in all aspects.
Comparing with Ostrich on this benchmark is problematic because it mostly fails due to unsupported syntactic features.  
\Trauplusspace is better than \Trau on all three benchmark sets. This shows that our proposed procedure is efficient in solving not only transducer examples, but also in transducer-less examples. 

To summarise our experimental results, we can see that:

\begin{itemize}
\item    \Trauplusspace handles more transducer examples in an efficient manner. This is illustrated by the Chain-Free and Slog suites. The experiment results on these benchmarks  show that \Trauplusspace outperforms \Traunospace. Many tests on which \Trau returns ``unknown'' are now successfully handled by \Trauplusspace.

\item  \Trauplusspace performs as well as existing tools on transducer-less benchmarks and in fact sometimes \Trauplusspace outperforms them. This is illustrated by the PyEx suite. In fact, this benchmark is handled very well by \Traunospace, but nevertheless, as evident from the table, our tool is doing better than \Traunospace. In fact, As \Trauplusspace out-performs \Trau in some examples in the PyEx suite. In those examples, \Trau returned ``unknown' while \Trauplusspace returned  ``unsat''. This means that the new over-approximation not only improves \Trau in transducer benchmarks, but it also improves \Trau in transducer-less examples. Furthermore, observe that the PyEx suite has only around 4000 ``unsat'' cases out of 25k cases. 

\end{itemize}

%\section{Extra Experiments}
\subsection{Additional Experiment}
\lukas{this section was in appendix as extra experiments. Include? (a better way to present that "Additional Experiment?"}
\label{section:extra_experiments}

\newcommand{\extratable}{
\newcolumntype{g}{>{\columncolor{white!20}}l}
\renewcommand{\arraystretch}{0.8}
\setlength{\tabcolsep}{3pt}
%\begin{table}[tbhp]
\begin{table}[t]
\centering
%\rowcolors[]{2}{gray!20}{white}
%\normalsize{
\begin{tabular}{|c|l|>{\centering\arraybackslash}p{1.5cm}|>{\centering\arraybackslash}p{1.5cm}|>{\centering\arraybackslash}p{1.5cm}|>{\centering\arraybackslash}p{1.5cm}|>{\centering\arraybackslash}p{1.5cm}|>{\centering\arraybackslash}p{1.5cm}|}
\hline
\rowcolor{white}						&				& Ostrich			& Z3-str3	& CVC4		& {\sc Trau} 	& \Trauplus \\ \hline	

\rowcolor{gray!20}\cellcolor{white}		&	sat        	& 28523				& 34495		& 35235			& 35246     	& 35246 \\ 
\rowcolor{gray!20}\cellcolor{white}		&	unsat		& 9872      			& 11799		& 12014 		& 12014      	& 12014 \\ 
										&	timeout     & 0		            & 350		& 35  			& 6	     		& 6 \\  
\multirow{-4}{*}{\parbox{15mm}{{\bf Kaluza} \\[-0mm] (47284)}}	& error/unkn. & 8889              & 640		& 0  			& 0	   			& 0 \\ \hline  
\rowcolor{gray!20}\cellcolor{white}		&	sat        	& 2             	& 8			& 8				& 8     	 	& 8\\ 
\rowcolor{gray!20}\cellcolor{white}		&	unsat		& 0             	& 0			& 0     		& 0     	 	& 0\\ 
										&	timeout     & 0              	& 0			& 0         	& 0	    		& 0\\  
\multirow{-4}{*}{\parbox{16mm}{{\bf Appscan}\\[-0mm] (8)}}		& error/unkn. & 6              & 0			& 0           	& 0	     	 	& 0\\ \hline  
\rowcolor{gray!20}\cellcolor{white}		&	sat        	& 0		        	& 8			& 8				& 8      		& 8\\ 
\rowcolor{gray!20}\cellcolor{white}		&	unsat		& 0           	  	& 4			& 4     		& 4     	 	& 4\\ 
										&	timeout     & 0              	& 0			& 0          	& 0	    		& 0\\  
\multirow{-4}{*}{\parbox{15mm}{{\bf PISA}\\[-0mm] (12)}}		& error/unkn. & 12            		& 0			& 0           	& 0	     		& 0\\ \hline \hline  
\rowcolor{gray!20}\cellcolor{white}		&	solved      	& 39387            	& 46314		& 47269			& 47298     	& 47298\\ 
\multirow{-2}{*}{\parbox{15mm}{{\bf Total}\\[-0mm] (47304)}}	& 	unsolved 	& 8907           & 990		& 35           	& 6	     	 	& 6\\ \hline   
\end{tabular}
%}

\caption{
\normalsize{Results of running each solver over Kaluza, Appscan, and PISA suites.  }
}
%\vspace{-0.4cm}
\label{fig:extra_results} 
\end{table}
}

In this section, we carry out additional experiments o
to ensure that our new decision procedure does not have negative effect on  \Trauplus.
We use benchmark suites such as Kaluza, AppScan, and PISA hat were executed successfully by \Trau and CVC4 previously. 
The Table \ref{fig:extra_results} shows that \Trauplusspace is also successful in solving these benchmarks. 

\medskip

%\noindent
{\it Kaluza suite.} The Kaluza suite \cite{Saxena10:kaluza} is generated by a JavaScript symbolic execution engine. It consists of 47284 test cases, including length, regular and (dis)equality constraints. For this suite, Ostrich returns unknown for 8889 cases. We contacted Ostrich' authors and received the explanation that those cases fall out from straight line fragment. CVC4 times out on 35 cases. Z3-str3 times out on 350 cases and cannot answer on 640 cases. \Trau and \Trauplusspace have the same performance, which is better than the other solvers as they time out  only in 6 cases.  
\medskip

%\noindent
{\it PISA and AppScan suite.} The PISA suite includes constraints from real-world Java sanitizer methods that were used in the evaluation of the PISA system \cite{Tateishi:2013:PIS:2522920.2522926}. The suite has 12 tests, including transducer constraints such as Substring, IndexOf, and Replace operations. The AppScan suite is derived from security warnings output by IBM Security AppScan Source Edition \cite{AppScan}. The suite has 8 tests, including  transducer constraints and (dis)equality constraints. In both suites, the performance of CVC4, Z3-str3, \Trau, and \Trauplusspace are able to solve all test cases.   

%	{\it PISA and AppScan suite.} The PISA suite includes constraints from real-world Java sanitizer methods that were used in the evaluation of the PISA system \cite{Tateishi:2013:PIS:2522920.2522926}. The suite has 12 tests, including transducer constraints such as Substring, IndexOf, and Replace, and (dis)equality constraints. The AppScan suite is derived from security warnings output by IBM Security AppScan Source Edition \cite{AppScan}. The suite has 8 tests, including regular membership constraints, transducer constraints and (dis)equality constraints.
%	The experimental results are summarised in Table~\ref{fig:kaluzaresults}. In both suites, the performance of  \Trau is comparable to Z3-str3 (they  are able to solve all test cases). CVC4 times out on 1 test case in each suite.  {\sc Trau} cannot run these two suites since it does not support transducer constraints. 

\extratable

\section{Related Work}

Already in  1946, Quine \cite{Quine46} showed that the first order theory
of string equations is undecidable.
An important line of work has been to identify subclasses
for which decidability can be achieved.
The pioneering work by Makanin \cite{makanin} proposed a decision
procedure for quantifier-free word equations, i.e., Boolean combinations of
equalities and disequalities, where the variables may denote words of 
arbitrary lengths.
The decidability and complexity of different subclasses
have been considered by several works, e.g.
\cite{Plandowski99,Plandowski06,Matiyasevich08,Robson90,Schulz90,Ganesh13decide,DBLP:journals/corr/GaneshB16}.
Generalizations of the work of Makanin by adding
new types of constraints have been difficult to achieve.
For instance, the satisfiability of word equations combined with length
constraints of the form $\sizeof\xvar=\sizeof\yvar$ is open
\cite{buchi:definability}.
Recently, 
regular and especially relational transducers constraints were identified
as a strongly desirable feature of string languages 
especially in the context software analysis with an emphasis on security. 
Adding these to the mix leads immediately to undecidability \cite{morvan}
and hence numerous decidable fragments were proposed \cite{string14,BFL13,BL16,Chen:2018,Chen:2019}. 
From these, the straight line fragment of \cite{BL16} is the most general decidable combination of concatenation and transducers. It is however incomparable to the acyclic fragment of \cite{string14} (which does not have transducers but could be extended with them in a straightforward manner).
Some works add also other syntactic features, such as \cite{Chen:2018,Chen:2019}, 
but the limit of decidable combinations of the core string features---transducers/regular constraints, length constraints, and concatenation stays at \cite{BL16} and \cite{string14}.
The weakly chaining decidable fragment present in this paper significantly generalises both these fragments in a practically relevant direction.

The strong practical motivation in string solving led to a rise of a number of SMT solvers that do not always provide completeness guarantees but concentrate on solving practical problem instances, 
through applying a variety of calculi and algorithms.
A number of tools handle string constraints by means of {\it length-based under-approximation} and
translation to
bit-vectors~\cite{Kiezun09hampi,Saxena10:kaluza,saxena:flax}, assuming
a fixed upper bound on the length of the possible solutions. 
Our method on the other hand allows to analyse constraints without a length limit and with completeness guarantees.
More recently, also {\it DPLL(T)-based} string solvers lift the
restriction of strings of bounded length; this generation of solvers
includes Z3-str3 \cite{Berzish2017Z3str3AS}, Z3-str2~\cite{Zheng13z3str}, 
CVC4~\cite{LiaEtAl-CAV-14}, S3P~\cite{trinh2014:s3,Trinh2016},
Norn~\cite{norn15}, 
Trau~\cite{trau18}, 
Sloth~\cite{sloth}, 
and 
Ostrich~\cite{Chen:2019}. DPLL(T)-based solvers
handle a variety of string constraints, including word equations, regular expression membership, length constraints, and (more rarely)
regular/rational relations; the solvers are not complete for the full
combination of those constraints though, and often only decide a (more
or less well-defined) fragment of the individual constraints. 
Equality constraints are normally handled by means of splitting into simpler
sub-cases, in combination with powerful techniques for Boolean
reasoning to curb the resulting exponential search space. 
Our implementation is combining strong completeness guarantees of \cite{sloth} extended to handle the fragment proposed in this paper with an efficient approximation techniques of \cite{trau18} 
and its performance on existing benchmarks compares favourably with the most efficient of the above tools.

A further direction is {\it automata-based} solvers for analyzing
string-manipulated programs. Stranger~\cite{fangyu:stranger} soundly
over-approximates string constraints using regular languages, and
outperforms DPLL(T)-based solvers when checking single execution
traces, according to some evaluations~\cite{KauslerS14}.  It has
recently also been observed~\cite{DBLP:conf/cav/WangTLYJ16,sloth} that
automata-based algorithms can be combined with model checking
algorithms, in particular IC3/PDR, for more efficient checking of the emptiness for automata. However, many kinds of constraints, including length
constraints and word equations, cannot be
handled by automata-based solvers in a complete manner.  

\section{Conclusions}
We have proposed a decision procedure for the so far largest fragment of string constraints which combine concatenation with transducer constraints and legth constraints.
We call the fragment chain-free. We also propose an extension of the chain-free string constraints to weakly chaining constraints, that can be tansformed to the chain-free form by a preprocessing step.
We have also proved that the chain-free constraints are in a certain sense the limit of what can be decied about general transducer constraints combined with word equations.
We have implemented the new decision procedures in the tool \Trauplus\ and experimentally demonstarted its practical relevance in a comparison with other string solvers.
One of the possible future directions is to work towards improving the efficiency of solving the automata problems that are in the core of the decision procedure, related to the combination of regular/transducer constraints and concatenation,
and continue our efforts in develplment of a mature string solver.

\section*{Acknowledgements}\lukas{double check}
This work has been supported by the Czech Ministry of Education and Health, (project No. 19-24397S). 

\bibliographystyle{splncs04}

\bibliography{references}

%\vfill

%\eject
%\appendix
%\input{sections/experiments_extra}

\end{document}